\documentclass[journal=jctcce,manuscript=article]{achemso}

\usepackage[version=3]{mhchem} 
\usepackage{subfiles}
\usepackage{amsmath}
\usepackage{amssymb}
\usepackage{physics}
\usepackage{graphicx}
\usepackage{dcolumn}
\usepackage{bm}
\usepackage{booktabs}
\usepackage{makecell}
\usepackage{subcaption}

\author{Kyle Bystrom}
\email{kylebystrom@g.harvard.edu}
\author{Stefano Falletta}
\email{stefanofalletta@g.harvard.edu}
\author{Boris Kozinsky}
\altaffiliation{Robert Bosch LLC Research and Technology Center, Cambridge, MA, USA}
\email{bkoz@seas.harvard.edu}
\affiliation{Harvard John A. Paulson School of Engineering and Applied Sciences}

\title{Addressing the Band Gap Problem with a Machine-Learned Exchange Functional}

\abbreviations{DFT,ML,KS,HF,XC}
\SectionNumbersOn

\begin{document}
\begin{singlespace} 
\begin{tocentry}

\includegraphics{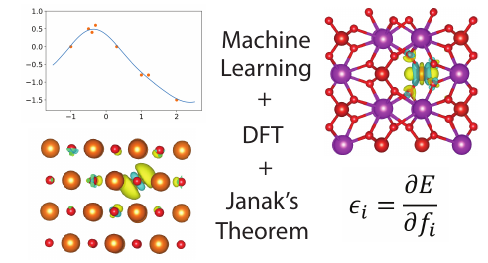}

\end{tocentry}

\begin{abstract}
    The systematic underestimation of band gaps is one of the most fundamental challenges in semilocal density functional theory (DFT). In addition to hindering the application of DFT to predicting electronic properties, the band gap problem is intimately related to self-interaction and delocalization errors, which make the study of charge transfer mechanisms with DFT difficult. In this work, we present two key innovations to address the band gap problem. First, we design an approach for machine learning density functionals based on Gaussian processes to explicitly fit single-particle energy levels. Second, we introduce novel nonlocal features of the density matrix that are expressive enough to fit these single-particle levels. Combining these developments, we train a machine-learned functional for the exact exchange energy that predicts molecular energy gaps and reaction energies of a wide range of molecules in excellent agreement with reference hybrid DFT calculations. In addition, while being trained solely on molecular data, our model predicts reasonable formation energies of polarons in solids, showcasing its transferability and robustness. Our approach generalizes straightforwardly to full exchange-correlation functionals, thus paving the way to the design of novel state-of-the-art functionals for the prediction of electronic properties of molecules and materials.
\end{abstract}

\section{Introduction}

The band gap problem, referring to the difficulty of predicting accurate fundamental band gaps with Kohn-Sham density functional theory (DFT)~\cite{kohn_self-consistent_1965,perdew_density_1986}, is one of the most critical challenges in modern electronic structure theory. The problem arises from an insufficient description of the exchange-correlation (XC) energy, particularly the self-interaction error \cite{perdew_density-functional_1982,ruzsinszky2007JCP,zhang1998JPC,yang2000PRL,mori2006JCP}. Specifically, semilocal DFT is notorious for systematically underestimating band gaps~\cite{perdew_density_1986}. A variety of approaches have been developed to address the band gap problem, most of which are based on hybrid DFT, in which a fraction of the exact exchange energy is mixed into a conventional semilocal XC functional~\cite{becke_densityfunctional_1993}. Hybrid DFT tends to predict more reasonable band gaps for molecules and solids~\cite{heyd_hybrid_2003,borlido_large-scale_2019,miceli2018PRB,yang2022JPCL,yang2023npj}, though obtaining consistently accurate band gaps requires system-specific tuning of the fraction of exact exchange due to different levels of screening in different materials~\cite{stein_fundamental_2010}. (For simplicity, we also refer to molecular HOMO-LUMO gaps as band gaps.) Significant research has been dedicated to addressing this problem nonempirically. For example, optimally tuned hybrid functionals and related methods~\cite{wing_band_2021,miceli2018PRB,yang2022JPCL,yang2023npj} tune the fraction of exact exchange based on exact conditions given by Janak's theorem~\cite{janak_proof_1978}. Other approaches include local~\cite{Borlido2018JCTC} and dielectric-dependent~\cite{Zheng2019PRM} hybrid functionals, which can be employed to study heterogeneous systems, where the ideal fraction of exact exchange is not constant in real-space. Machine learning-based methods to tune the parameter empirically for each chemical system have also been developed~\cite{Ju2021JPCL,Ju2024JPCA}. Another approach to acquiring accurate band gap predictions, called Koopmans functionals~\cite{Dabo2010,Nguyen2018}, uses orbital-dependent (but not exact exchange-based) information to enforce piecewise linearity of the energy with respect to the number of electrons, thereby removing self-interaction. While these hybrid DFT and orbital-dependent approaches tend to come at significantly higher computational cost than semilocal DFT, they demonstrate that addressing the band gap problem is possible, even for large and complex systems.

Concurrently with these developments, research on the the use of machine learning to improve the accuracy of XC functionals has gained traction over the past few years~\cite{snyder_finding_2012} and has shown significant promise for improving DFT predictions of molecular reaction energies and even strongly correlated systems~\cite{kirkpatrick_pushing_2021}. Approaches for training the XC functional range from straightforwardly fitting energy densities from higher-level theories~\cite{lei_design_2019,bystrom_cider_2022} to using fully differentiable DFT simulations to train functionals self-consistently~\cite{li_kohn-sham_2021,kasim_learning_2021,dick_highly_2021,chen_deepks_2021}. Likewise, choices of model inputs range from conventional DFT ingredients from Jacob's ladder~\cite{perdew_jacobs_2001,kirkpatrick_pushing_2021,dick_highly_2021,nagai_machine-learning-based_2022} to novel geometry-independent~\cite{lei_design_2019,nagai_completing_2020,bystrom_cider_2022,bystrom_nonlocal_2023,sahoo_self-consistent_2024,riemelmoser_machine_2023} and geometry-dependent~\cite{chen_deepks_2021,dick_machine_2020,chen_development_2024} featurizations of the density. Amid all this variety, however, the focus of previous work has been on training the machine learning model to total and relative energies of chemical systems. This does not guarantee the accurate learning of band gaps. Since the fundamental band gap in Kohn-Sham DFT is related to the derivative of the energy with respect to the number of electrons,~\cite{janak_proof_1978} it necessarily depends on the behavior of the functional for a fractional number of electrons. Good proxies for these fractional electron systems are not common in molecular benchmark datasets, so it is expected that obtaining accurate gaps will require dedicated training. 

Accurately fitting energy gaps could be useful on multiple fronts. First off, accurate gaps are valuable in and of themselves for predicting the electronic properties of materials. Second, errors in energy gaps are closely tied to self-interaction and delocalization error~\cite{cohen_insights_2008}, so fitting energy gaps could potentially help improve the accuracy of DFT densities and barrier heights, which are currently hindered by self-interaction in conventional DFT. An initial illustration of this was provided by the DM21 functional~\cite{kirkpatrick_pushing_2021}, which achieved state-of-the-art performance on molecular benchmarks partly by explicitly fitting systems with fractional charge. The behavior of a given functional for fractionally charged systems is closely related to its predictions of energy gaps~\cite{cohen_insights_2008}, so explicit fitting of energy gaps could help generalize and expand the basic ideas that made DM21 effective.

In this work, to address the band gap problem in the context of machine-learned functionals, we introduce a methodology for explicitly learning their contributions to both the total energies of chemical systems and the single-particle energy levels of those systems. This enables the accurate predictions of energy gaps, since these are calculated as differences of highest-occupied and lowest-unoccupied energy levels. To approach this problem, we build on top of the CIDER framework~\cite{bystrom_cider_2022,bystrom_nonlocal_2023} for designing machined-learned functionals. CIDER has several key components that make it an ideal starting point for new developments. First, CIDER uses features of the charge density that are both nonlocal and computationally efficient to evaluate, resulting in both accurate and computationally efficient models. These features are also designed to strictly enforce the uniform scaling rule of the exchange energy~\cite{levy_hellmann-feynman_1985}, and exact constraints can improve the predictive power and transferability of XC functionals~\cite{kaplan_predictive_2023}. Second, CIDER uses carefully tuned Gaussian process models to produce functional forms that are both highly flexible and numerically stable, enabling practical simulations of hundreds of atoms for complex systems like point defect structures~\cite{bystrom_nonlocal_2023}.

Starting from this previous work, we adapt the CIDER framework to fit training data composed of both energy and single-particle energy levels with a single model (as opposed to energy only). Next, we introduce novel features of the density matrix that are sufficiently expressive to capture both the exchange energy and its contributions to the single-particle energy levels. Finally, we combine these two developments to train an exchange functional and show that it simultaneously provides a good description of the relative energies and energy gaps of molecular systems. While our functional is trained solely on isolated molecular systems, we show that it can be effectively used to localize polarons in solids and predict their formation energies by reducing the electron self-interaction, which is a promising illustration of the transferability of the method and of its potential to describe solid-state systems. While the current model operates as a surrogate for hybrid DFT and therefore inherits its limitations, such as the need for system-specific tuning and poor description of strong correlation, it can be straightforwardly extended to full XC functionals. In addition, the new model circumvents the need to evaluate the exact exchange energy used in hybrid DFT, which could enable the development of lower-cost alternatives to the state-of-the-art hybrid DFT methods used for band gap prediction.

This work is an extension of our recent formulation~\cite{bystrom_nonlocal_2023} of CIDER functionals for efficient and scalable production-level calculations, which included a full implementation of analytical force and stress evaluation. Here we introduce and test novel features based on the density matrix and a new model training approach, laying the foundations for future work on performance optimization and force/stress implementation. Our new models are already sufficiently well-optimized to perform 100-atom isolated or periodic SCF calculations on a few dozen CPU cores, but as initially implemented do not always provide a significant performance gain over hybrid DFT at this system size. However, due to the more favorable computational cost scaling of the new features with system size compared to hybrid DFT, we expect that optimizations of this new approach will be useful for large-scale materials simulations.

The rest of the paper is organized as follows. Section~\ref{sec:theory} covers the modified Gaussian process model used to simultaneously fit energy and single-particle energy levels, and it describes the new features used as input to the model. Section~\ref{sec:methods} provides computational details for the training and benchmarking calculations performed. Section~\ref{sec:results} presents and discusses the accuracy of the new functional for predicting molecular properties, energy gaps, and polaron formation energies. Finally, Section~\ref{sec:conclusion} concludes and discusses future directions and extensions of this new methodology.

\section{Theory}~\label{sec:theory}

This section is divided into three parts. Section~\ref{sec:theory_overview} introduces the basic definitions and concepts that are critical to understanding the band gap problem and fitting single-particle energies. Section~\ref{sec:theory_gp} describes how Gaussian process regression can be used to fit XC functionals to the energies of chemical systems as well as their single-particle energy levels. Section~\ref{sec:theory_sdmx} describes the new features we use to train the functional in this work.

\subsection{Janak's Theorem and Derivative Discontinuity}\label{sec:theory_overview}

In Kohn-Sham DFT~\cite{kohn_self-consistent_1965,seidl_generalized_1996,gorling_hybrid_1997}, the one-body reduced density matrix $n_1(\mathbf{r}, \mathbf{r}')$ and the electron density $n(\mathbf{r})$ are written in terms of a set of fictitious orbitals $\phi_i(\mathbf{r})$ and their occupation numbers $f_i$ as
\begin{align}
    n_1(\mathbf{r}, \mathbf{r}') &= \sum_i f_i \phi_i^*(\mathbf{r}) \phi_i(\mathbf{r}') \\
    n(\mathbf{r}) &= n_1(\mathbf{r},\mathbf{r})
\end{align}
where the index $i$ labels the electron states. All of our models are implemented for both the spin-restricted formalism and the (collinear) spin-unrestricted formalism, for which the density matrix also has a spin index $\sigma$:
\begin{equation}
    n_{1,\sigma}(\mathbf{r},\mathbf{r}') = \sum_i f_{\sigma,i}\phi_{\sigma,i}^*(\mathbf{r})\phi_{\sigma,i}(\mathbf{r}')
\end{equation}
For simplicity, for most of this work we will use the spin-restricted formalism with $f_i\in[0,2]$.

Within the spin-restricted formalism, the total energy of the system is written as a functional of $n_1$
\begin{align}
    E[n_1] =& -\frac{1}{2} \int \dd[3]\mathbf{r} \nabla^2_\mathbf{r'}n_1(\mathbf{r}, \mathbf{r}')\rvert_{\mathbf{r}'=\mathbf{r}}\notag\\
    &+ \int \dd[3]\mathbf{r}\, v_\text{ext}(\mathbf{r}) n(\mathbf{r})\notag\\
    &+ \frac{1}{2} \int \dd[3]\mathbf{r}\, \dd[3]\mathbf{r}'\, \frac{n(\mathbf{r}) n(\mathbf{r}')}{|\mathbf{r}-\mathbf{r}'|} + E_\text{xc}[n_1] \label{eq:etot_gks}
\end{align}
where $v_\text{ext}(\mathbf{r})$ is the external potential, and $E_\text{xc}[n_1]$ is the exchange-correlation functional. Finding the ground-state energy requires minimizing eq~\ref{eq:etot_gks}. To do so, one must self-consistently solve for the orbitals $\phi_i(\mathbf{r})$ using the partial differential equation
\begin{equation}
    -\frac{1}{2}\nabla^2\phi_i(\mathbf{r}) + \int \dd[3]\mathbf{r}'\, v_\text{eff}[n_1](\mathbf{r},\mathbf{r}')\phi_i(\mathbf{r}')=\epsilon_i\phi_i(\mathbf{r}) \label{eq:effective_ham}
\end{equation}
with the nonlocal effective potential
\begin{align}
    v_\text{eff}[n_1](\mathbf{r}, \mathbf{r}') =& \left[v_\text{ext}(\mathbf{r}) + \int \dd[3]\mathbf{r}'' \frac{n(\mathbf{r}'')}{|\mathbf{r}-\mathbf{r}''|} \right] \delta(\mathbf{r}-\mathbf{r}') \notag\\
    &+ \fdv{E_\text{xc}[n_1]}{n_1(\mathbf{r}, \mathbf{r}')}
\end{align}
The eigenvalues $\epsilon_i$ are the DFT single-particle energy levels. In this work, we employ generalized Kohn-Sham DFT~\cite{seidl_generalized_1996,gorling_hybrid_1997} (the typical formalism used for meta-GGA~\cite{sun_strongly_2015} and hybrid functionals~\cite{becke_densityfunctional_1993}), in which the XC energy $E_\text{xc}[n_1]$ is an explicit functional of the density matrix rather than the density. The reason for this choice is that the exact XC energy has a discontinuous derivative with respect to the number of electrons $N$~\cite{perdew_physical_1983,mori-sanchez_derivative_2014}:
\begin{align}
    \Delta_\text{xc}(N)&=\pdv{E_\text{xc}(N)}{N_+} - \pdv{E_\text{xc}(N)}{N_-} \\
    \Delta_\text{xc}(N)&\ne 0\,\,\,\,\,\,\,\,\,\,\,\,N\in\mathbb{Z}
\end{align}
where $\pdv{}{N_-}$ and $\pdv{}{N_+}$ are the left and right-hand derivatives with respect to $N$, respectively. This derivative discontinuity $\Delta_\text{xc}(N)$ cannot be captured by a smooth, explicit functional of the density $E_\text{xc}[n]$, but it can be described by a density matrix functional $E_\text{xc}[n_1]$~\cite{perdew_understanding_2017,yang_more_2016}. Accurately predicting $\Delta_\text{xc}(N)$ is critical for determining the ionization potential (IP), electron affinity (EA), and band gap (BG) of a system because these quantities depend on the derivative of the total energy with respect to $N$:
\begin{align}
    \text{IP} &= -\pdv{E(N)}{N_-}\\
    \text{EA} &= \pdv{E(N)}{N_+}\\
    \text{BG} &= \text{EA} + \text{IP} = \pdv{E(N)}{N_+}-\pdv{E(N)}{N_-}
\end{align}
The question is then how to practically compute terms like $\pdv{E(N)}{N_-}$. Thankfully, Janak's theorem~\cite{janak_proof_1978} allows us to write the derivatives of the energy with respect to orbital occupation numbers $f_i$ and eigenvalues $\epsilon_i$ as
\begin{equation}
    \epsilon_i = \pdv{E[n_1]}{f_i} \label{eq:eigval_pdv_energy}
\end{equation}
This theorem allows us to write the IP, EA, and BG as
\begin{align}
    \text{IP} &= -\pdv{E[n_1]}{f_\text{VB}} = -\epsilon_\text{VB}\label{eq:ip_occ_deriv}\\
    \text{EA} &= \pdv{E[n_1]}{f_\text{CB}} = \epsilon_\text{CB}\label{eq:ea_occ_deriv}\\
    \text{BG} &= \pdv{E[n_1]}{f_\text{CB}}-\pdv{E[n_1]}{f_\text{VB}}=\epsilon_\text{CB}-\epsilon_\text{VB}\label{eq:bg_occ_deriv}
\end{align}
In the above equation, we denote the eigenvalue corresponding to the valence band maximum (for solids) or highest-energy occupied orbital (for molecules) as $\epsilon_\text{VB}$, and the eigenvalue corresponding to the conduction band minimum or lowest-energy unoccupied orbital as $\epsilon_\text{CB}$. The corresponding occupations of these orbitals will be denoted $f_\text{VB}$ and $f_\text{CB}$, respectively.

Working from eqs~\ref{eq:ip_occ_deriv}--\ref{eq:bg_occ_deriv}, there are two key problems to solve in order to learn an XC functional $E_\text{xc}[n_1]$ that accurately predicts the band gap and related properties. First, we must design a way to train the derivatives of our functional with respect to orbital occupation numbers to reference band gaps, ionization potentials, and electron affinities. We address this problem in Section~\ref{sec:theory_gp}. Second, we must design features with explicit density matrix dependence and significant derivative discontinuities, or else it will be implausible to design a functional obeying the above relationships (since pure density functionals lack a derivative discontinuity~\cite{perdew_understanding_2017,yang_more_2016}). We address this challenge in Section~\ref{sec:theory_sdmx}.

Before proceeding, we note that $E_\text{xc}[n_1]$ can be decomposed into exchange and correlation parts, the former of which can be written exactly in terms of the spin-polarized density matrix:
\begin{align}
    E_\text{xc}[n_1] &= E_\text{x}[n_1] + E_\text{c}[n_1] \\
    E_\text{x}^\text{exact}[n_1] &= -\frac{1}{2} \sum_\sigma \int \dd[3]\mathbf{r} \int \dd[3]\mathbf{r}' \frac{|n_{1,\sigma}(\mathbf{r},\mathbf{r}')|^2}{|\mathbf{r}-\mathbf{r}'|}\label{eq:exact_exchange_energy}
\end{align}
where $E_\text{x}$ is the exchange energy, and $E_\text{c}$ is the correlation energy. Equation~\ref{eq:exact_exchange_energy} is explicitly known but computationally expensive to evaluate, and it is used within hybrid DFT to obtain much more accurate band gaps than those obtained with semilocal DFT. For these reasons, it is an excellent learning target for testing our new methodology, and having a good exact exchange ``surrogate'' is potentially useful in its own right for accelerating hybrid DFT calculations. Therefore, in this work, rather than learning the full XC functional, we learn $E_\text{x}^\text{exact}[n_1]$ and substitute it in place of exact exchange in existing hybrid functionals.

\subsection{Gaussian Process Models for DFT Energies and Eigenvalues\label{sec:theory_gp}}

In this section, we introduce our Gaussian process-based approach for fitting exchange and correlation functionals to both total energies and single-particle energy levels. Our current implementation only addresses the exchange functional, but the approach described in this section generalizes to the full XC functional as well. This methodology builds on the Gaussian process approach from our previous work for fitting total exchange energies~\cite{bystrom_nonlocal_2023}, but we describe the full approach in this section for sake of clarity and generality.

\subsubsection{Background}

We briefly discuss the components relevant to fit and evaluate a Gaussian process predictive mean, i.e.\ the function that the Gaussian process model learns. The detailed theory of Gaussian process regression, which is a Bayesian statistical learning method, can be found in the textbook by Rasmussen and Williams~\cite{rasmussen_gaussian_2006} (Chapter 2).

Consider the problem of learning some function $f(\mathbf{x})$, with $\mathbf{x}$ being a feature vector of independent variables. Let $\mathbf{X}$ be a feature matrix, where each row $\mathbf{x}_i$ is the feature vector for training point $i$, and let $\mathbf{y}$ be a target vector, where each element $y_i$ is the observed value of the target function for training point $i$. Then, there are two key components necessary to construct a Gaussian process regression for $f(\mathbf{x})$. The first is the covariance kernel $k(\mathbf{x}, \mathbf{x}')$, which defines the covariance between the values of the predictive function for two points, i.e. $k(\mathbf{x}, \mathbf{x}')=\text{Cov}(f(\mathbf{x}), f(\mathbf{x}'))$. The matrix of covariances between training points is denoted $\mathbf{K}$ with matrix elements $K_{ij}=k(\mathbf{x}_i, \mathbf{x}_j)$. The second is the noise labels $\sigma_i$, indicating the estimated prior uncertainty on each training observation $i$. We will denote the diagonal matrix whose entries are the noise covariances ${\sigma_i^{}}^2$ as $\Sigma_\text{noise}$. Using these components, the predictive function for a test point $\mathbf{x}_*$ takes the form~\cite{rasmussen_gaussian_2006}
\begin{equation}
    f(\mathbf{x}_*) = \sum_\alpha k(\mathbf{x}_*, \mathbf{x}_i) \alpha_i
\end{equation}
with the following definition for the weight vector $\boldsymbol{\alpha}$:
\begin{equation}
    \boldsymbol{\alpha} = \left(\mathbf{K} + \Sigma_\text{noise}\right)^{-1} \mathbf{y}
\end{equation}
In the case of fitting energies of molecules and solids, we need to fit an extensive quantity in which the training labels are integrals of the predictive function $f(\mathbf{x})$ over real space. The next section covers adjustments to the Gaussian process model necessary to perform this task.

\subsubsection{Fitting Total Energy Data} \label{sec:fit_energy}

This section provides a more general description of the approach for fitting density functionals that was presented in our previous work~\cite{bystrom_nonlocal_2023}. Consider an extensive quantity $F$, which is a contribution to the total electronic energy of a chemical system. While our approach currently focuses on the exchange energy $E_\text{x}$, it can also be used to fit the correlation energy, full XC energy, or other extensive quantities. We can fit $F$ by learning a function that gets integrated over real-space to yield $F$ for a given system:
\begin{equation}
    F = \int \dd[3]\mathbf{r}\,f\left(\mathbf{x}(\mathbf{r})\right) \label{eq:F_int}
\end{equation}
In the above equation, $f\left(\mathbf{x}(\mathbf{r})\right)$ is the predictive function for the energy density to be learned by the Gaussian process. In practice, the integral in eq~\ref{eq:F_int} must be performed numerically. For a given chemical system indexed by $m$, we write
\begin{equation}
    F^m = \sum_{g\in m} w_g^m f\left(\mathbf{x}_g^m\right) \label{eq:extensive_functional}
\end{equation}
where $g$ indexes quadrature points and $w_g^m$ are the respective quadrature weights.

In order to construct a Gaussian process for eq~\ref{eq:extensive_functional}, we need to compute the covariance between two observed values $F^m$ and $F^n$ for chemical systems $m$ and $n$. Because the covariance between sums of random variables can be written as
\begin{align}
    \text{Cov}(a+b,c+d)=\ &\text{Cov}(a,c)+\text{Cov}(b,c)\notag\\&+\text{Cov}(a,d)+\text{Cov}(b,d) \label{eq:cov_of_sum}
\end{align}
the covariances between the numerical integrals $F^m$ and $F^n$ can be written in terms of the covariance kernel for $f(\mathbf{x})$ as
\begin{align}
    \text{Cov}(F^m, F^n) =& \sum_{g \in m} \sum_{h \in n} w_g^m w_h^n \text{Cov}(f(\mathbf{x}_g^m), f(\mathbf{x}_h^n)) \\
    =& \sum_{g \in m} \sum_{h \in n} w_g^m w_h^n k(\mathbf{x}_g^m, \mathbf{x}_h^m) \label{eq:exc_cov_exact}
\end{align}
where $k(\mathbf{x}, \mathbf{x}')$ is the covariance kernel for $f(\mathbf{x})$. Computing the above double numerical integral directly would be expensive. To overcome this issue, we define a small set of ``control points'' $\tilde{\mathbf{x}}_a$ and approximate $\text{Cov}(F^m, F^n)$ using a resolution-of-the-identity approximation:
\begin{align}
    \text{Cov}(F^m, F^n) = K_{mn} &\approx \tilde{\mathbf{k}}_m \tilde{\mathbf{K}}^{-1} \tilde{\mathbf{k}}_n \label{eq:roi_cov} \\
    \left(\tilde{\mathbf{K}}\right)_{ab} &= k(\tilde{\mathbf{x}}_a, \tilde{\mathbf{x}}_b) \\
    \left(\tilde{\mathbf{k}}_m\right)_a &= \sum_{g\in m} w_g^m k(\mathbf{x}_g^m, \tilde{\mathbf{x}}_a) \label{eq:roi_cov_component}
\end{align}
Using this definition of the covariance kernel, the predictive function can be expressed as
\begin{align}
    f(\mathbf{x}_*) &= \sum_a k(\mathbf{x}_*, \mathbf{\tilde{x}}_a) \alpha_a \label{eq:gp_sum_formula} \\
    \boldsymbol{\alpha} &= \sum_m \mathbf{\tilde{k}}_m \left\{\left[\mathbf{K} + \Sigma_{\text{noise}}\right]^{-1} \mathbf{y}\right\}_m \label{eq:gp_predictive}
\end{align}
with $\mathbf{K}$ being the matrix with elements defined in eq~\ref{eq:roi_cov}, $\mathbf{\tilde{k}}_m$ being the vector defined in eq~\ref{eq:roi_cov_component}, and $\mathbf{y}$ being the vector of training labels $F^m$.

\subsubsection{Fitting Single-Particle Energy Levels} \label{sec:fit_eigvals}

Having summarized the Gaussian process regression scheme for total energies, we now extend it to fitting single-particle energy levels. From eq~\ref{eq:eigval_pdv_energy}, $\epsilon_i^m$ is the partial derivative of the total energy of system $m$ with respect to the occupation number $f_i^m$ of orbital $i$. By inserting eq~\ref{eq:eigval_pdv_energy} into eq~\ref{eq:extensive_functional}, one can write the eigenvalues $\epsilon_i^m$ (or contributions to the eigenvalues) in terms of the predictive function $f(\mathbf{x})$ for the total energy (or contributions to the total energy):
\begin{align}
    \epsilon_i^m &= \sum_{g \in m} w_g^m \pdv{f(\mathbf{x}_g^m)}{f_i^m} \\
    \epsilon_i^m &= \sum_{g \in m} w_g^m \pdv{f(\mathbf{x}_g^m)}{\mathbf{x}_g^m} \cdot \pdv{\mathbf{x}_g^m}{f_i^m} \label{eq:eigval_feat_grad}
\end{align}
One can then write the covariance between a given energy level $\epsilon_i^m$ in chemical system $m$ and the energy of another chemical system $F^n$ or another energy level $\epsilon_j^n$:
\begin{align}
    \text{Cov}\left(\epsilon_i^m, F^n\right)&=\text{Cov}\left(\pdv{F^m}{f_i^m}, F^n\right) = \tilde{\mathbf{d}}_{mi} \tilde{\mathbf{K}}^{-1} \tilde{\mathbf{k}}_n \\
    \text{Cov}\left(\epsilon_i^m, \epsilon_j^n\right)&=\text{Cov}\left(\pdv{F^m}{f_i^m}, \pdv{F^n}{f_j^n}\right) = \tilde{\mathbf{d}}_{mi} \tilde{\mathbf{K}}^{-1} \tilde{\mathbf{d}}_{nj} \\
    \left(\tilde{\mathbf{d}}_{mi}\right)_a &= \sum_{g\in m} w_g^m \pdv{\mathbf{x}_g^m}{f_i^m} \cdot \pdv{}{\mathbf{x}_g^m} k(\mathbf{x}_g^m, \tilde{\mathbf{x}}_a)
\end{align}
The above equations, like eq~\ref{eq:roi_cov} for energy data, allow eigenvalue data to be trained using Gaussian process regression. One can fit single-particle energy levels using exact constraints, such as piecewise linearity and quadraticity constraints as discussed below in Section~\ref{sec:fit_eigval_to_constraint}, and/or using reference data such as ionization potentials and band gaps, as discussed in Section~\ref{sec:fit_eigval_to_data}.

Before proceeding, we note that the XC energy is one of several terms in the DFT total energy in eq~\ref{eq:etot_gks}. However, the other terms can be computed explicitly from the orbitals or density matrix. Therefore, given a set of orbitals sufficiently close to the exact ones, any training label for the total energy $E$ can also be used as a training label for $E_\text{xc}$ by subtracting all energy contributions besides the $E_\text{xc}$ from the training label $E$. These reference orbitals can be obtained by different means, such as by computing them with an existing functional (e.g.\ Kirkpatrick \emph{et al.}~\cite{kirkpatrick_pushing_2021}) or computing them self-consistently while training (e.g.\ Dick and Fern\'andez-Serra~\cite{dick_highly_2021}). The same concept applies to the energy levels $\epsilon_i$, since by eq~\ref{eq:eigval_pdv_energy} we can subtract the occupation number derivatives of all other energy terms from $\epsilon_i$ to find the XC contribution to $\epsilon_i$. We make this note to emphasize that all expressions for total energy and energy level training data below also serve as XC energy training data.

One can also train the correlation energy in isolation (by subtracting the exact exchange energy of eq~\ref{eq:exact_exchange_energy} from $E_\text{xc}$) or the exchange energy in isolation (by computing eq~\ref{eq:exact_exchange_energy} for the training systems and taking its derivative with respect to orbital occupation to obtain the exchange contributions to the eigenvalues). In this work, we pursue the latter approach of training only the exchange energy to create surrogates for hybrid DFT.

\subsubsection{Fitting Single-Particle Energy Levels Using Piecewise Linearity and Quadraticity Constraints} \label{sec:fit_eigval_to_constraint}

The total energy $E$ of a system obeys a piecewise linearity constraint for fractional electron count $N$~\cite{perdew_density-functional_1982,cohen_fractional_2008,cohen_insights_2008}. The energy as a function of $N$ is
\begin{equation}
    E(N) = \left(1 - N + \lfloor N \rfloor\right) E\left(\lfloor N \rfloor\right) + \left(N - \lfloor N \rfloor\right) E\left(\lfloor N \rfloor + 1\right)
\end{equation}
where $\lfloor x \rfloor$ is the floor of $x$. For simplicity (and without loss of generality), we can write the expression above for $N\in[0,1]$:
\begin{equation}
    E(N) = (1-N)E(0) + NE(1) \label{eq:pwl}
\end{equation}
Because the eigenvalues $\epsilon_\text{VB}$ and $\epsilon_\text{CB}$ are the left-hand and right-hand derivatives of the energy with respect to particular number, respectively (as discussed in Section~\ref{sec:theory_overview}), they also have rigorous relationships with the total energy:
\begin{equation}
    E(1) - E(0) = \epsilon_\text{CB}(0) = \epsilon_\text{VB}(1)\label{eq:pwl_levels}
\end{equation}

Similarly, as derived in Appendix~\ref{app:quad}, the exchange energy is ``piecewise quadratic'' in the number of electrons under the frozen orbital approximation. By combining this quadratic form with the exchange contribution to the CB eigenvalue, $\epsilon_\text{x,CB}=\pdv{E_\text{x}}{f_\text{CB}}$, one can write the exchange energy for $N\in[0,1]$ as
\begin{equation}
    E_\text{x}(N) = (1-N)^2 E_\text{x}(0) + N^2 E_\text{x}(1) + \left(\epsilon_\text{x,CB}(0) + 2E_\text{x}(0) \right) N (1 - N)
\end{equation}
Equivalently,
\begin{equation}
    E_\text{x}(N) = (1-N)^2 E_\text{x}(0) + N^2 E_\text{x}(1) + \left(2E_\text{x}(1)-\epsilon_\text{x,VB}(1)\right) N (1 - N)
\end{equation}
Combining the above two equations,
\begin{equation}
    2(E_\text{x}(1)-E_\text{x}(0))=\epsilon_\text{x,CB}(0) + \epsilon_\text{x,VB}(1) \label{eq:pwq}
\end{equation}

The above conditions provide training data that does not require actually collecting any experimental or quantum chemical computational data. For example, one can fit to eq~\ref{eq:pwl_levels} by minimizing the loss function
\begin{align}
    \mathcal{L} =& \left(E^\text{model}(1) - E^\text{model}(0) - \epsilon_\text{CB}^\text{model}(0)\right)^2 \notag\\
    &+ \left(E^\text{model}(1) - E^\text{model}(0) - \epsilon_\text{VB}^\text{model}(1)\right)^2 \label{eq:pwl_loss}
\end{align}
where ``model'' indicates that the quantity is the value predicted by the machine learning model being trained. Note that in eq~\ref{eq:pwl_loss}, no reference values are required for the energy or energy levels; the training data arises from a physical requirement of the model rather than a computed or experimentally measured reference value. The loss function of eq~\ref{eq:pwl_loss} is closely analogous to how the actual Gaussian process training works, since in practice the Gaussian process minimizes a weighted mean-squared loss function subject to regularization.

One could also use the above strategy for the exchange energy via eq~\ref{eq:pwq}. However, one important caveat is that eq~\ref{eq:pwq} only applies under the frozen orbital approximation (c.f.\ Appendix~\ref{app:quad}), i.e. if there are no changes to the orbitals when going from $N=0$ to $N=1$. Since the frozen orbital approximation is not obeyed perfectly in real systems, there might be some (typically small) non-quadratic behavior in the exact exchange functional. Therefore, instead of training to the exact quadraticity in eq \ref{eq:pwq}, one can train to equivalence between exact exchange and model exchange:
\begin{align}
    2(&E_\text{x}^\text{model}(1)-E_\text{x}^\text{model}(0)) - \left(\epsilon_\text{x,CB}^\text{model}(0) + \epsilon_\text{x,VB}^\text{model}(1)\right) = \notag\\ & 2(E_\text{x}^\text{exact}(1)-E_\text{x}^\text{exact}(0)) - \left(\epsilon_\text{x,CB}^\text{exact}(0) + \epsilon_\text{x,VB}^\text{exact}(1)\right)\label{eq:quadraticity_constraint}
\end{align}
Even though the piecewise quadraticity of the exchange energy is only approximate due to the frozen orbital assumption, a perfect machine learning model for exact exchange will reproduce eq~\ref{eq:quadraticity_constraint}. Therefore, while eq~\ref{eq:quadraticity_constraint} actually does require computing exact exchange energy training data, fitting the model to minimize the error for this condition helps give the model the same amount of quadratic behavior as exact exchange and therefore serves a similar role as the piecewise linearity constraint on the total energy.

\subsubsection{Fitting Single-Particle Energy Levels to Reference Data} \label{sec:fit_eigval_to_data}

One can also train single-particle energy levels to explicit reference data. For example, for a given reference electron affinity EA for a system, one can fit the total energy and band edges as
\begin{equation}
    E(1) - E(0) = \epsilon_\text{CB}(0) = \epsilon_\text{VB}(1) = \text{EA}
\end{equation}
Likewise, for an ionization potential IP,
\begin{equation}
    E(-1) - E(0) = -\epsilon_\text{VB}(0) = -\epsilon_\text{CB}(-1) = \text{IP}
\end{equation}
For band gaps,
\begin{equation}
    \epsilon_\text{CB}(0) - \epsilon_\text{VB}(0) = \text{EA} + \text{IP}
\end{equation}
For exchange, one can fit the quadratic component of the fractional charge energy curve to that of exact exchange, i.e. adding the following condition as a target:
\begin{equation}
    \epsilon_\text{x,CB}^\text{model}(0) - \epsilon_\text{x,VB}^\text{model}(1) = \epsilon_\text{x,CB}^\text{exact}(0) - \epsilon_\text{x,VB}^\text{exact}(1)\label{eq:exx_quadraticity}
\end{equation}
In addition, one can simply train the contributions of the model exchange to the single-particle energy levels to match the exact exchange contributions. For example, for the band gap, one can add as a target the condition
\begin{equation}
    \epsilon_\text{x,CB}^\text{model}(0) - \epsilon_\text{x,VB}^\text{model}(0)=\epsilon_\text{x,CB}^\text{exact}(0) - \epsilon_\text{x,VB}^\text{exact}(0)
\end{equation}

\subsection{Smoothed Density Matrix Exchange, a New Type of Feature for Learning Density Functionals} \label{sec:theory_sdmx}

In this section, we introduce what we call smoothed density matrix exchange (SDMX) features, which are computationally efficient, nonlocal featurizations of the density matrix that are quadratic in orbital occupations. The basic idea of SDMX is to ``smooth'' out the density matrix around a reference point $\mathbf{r}$ and also project it onto low-order spherical harmonics (i.e. principle quantum numbers of $\ell=0$ and $\ell=1$). As a result, one can obtain very simple (but nonlocal) proxies for the shape of the exchange hole $|n_1(\mathbf{r},\mathbf{r}')|^2$, which can be used for fitting both the exchange and correlation energies. The formulas and details for the different SDMX versions are described below. These features take significant inspiration from the ``Rung 3.5'' functionals developed by Janesko \emph{et al.}~\cite{janesko_rung_2010,Janesko2013,Janesko2014,Janesko2018}. There are a few key differences, however. For example, the Rung 3.5 approach requires multiplying the exact density matrix by a model density matrix to make an approximate exchange hole, whereas the SDMX approach multiplies a ``smoothed'' density matrix by itself. As a result, the SDMX features have quadratic dependence on the density matrix.

One key consideration in the design of the SDMX features is their behavior under uniform scaling of the density, i.e.
\begin{equation}
    n_\lambda(\mathbf{r})=\lambda^3n(\lambda\mathbf{r})\label{eq:uniform_scaling}
\end{equation}
This manipulation is important because the exact exchange energy obeys the relationship~\cite{levy_hellmann-feynman_1985}
\begin{equation}
    E_\text{x}[n_\lambda]=\lambda E_\text{x}[n]\label{eq:exx_uscale}
\end{equation}
The correlation energy does not have such a simple behavior, but it does have the low and high-density limits $\lim_{\lambda\rightarrow 0} E_\text{c}[n_\lambda]=\lambda C_0[n]$ and $\lim_{\lambda\rightarrow\infty} E_\text{c}[n_\lambda]>-\infty$~\cite{levy_density-functional_1991,kaplan_predictive_2023}, where $C_0[n]$ is a density functional that does not depend on $\lambda$. Therefore, if one wishes to design an XC functional that rigorously matches these known behaviors under uniform scaling, it is critical that the features used as input to the functional have a simple behavior under uniform scaling. In keeping with this design objective, all SDMX features have power law behavior under uniform scaling, as discussed below.

The construction of the most basic SDMX feature starts with a smoothed, spherically averaged density matrix
\begin{equation}
    \rho^0(R; \mathbf{r}) = \int \dd[3]\mathbf{r}'\, h(|\mathbf{r}'-\mathbf{r}|; R) n_1(\mathbf{r}', \mathbf{r})
\end{equation}
where $R$ is a length-scale parameter, and $h$ is a smooth convolution kernel. Then, one can construct the set of features $H_j^0(\mathbf{r})$ as
\begin{equation}
    H_j^0(\mathbf{r}) = 4\pi \int \dd R\, R^{2-j} \left|\rho^0(R; \mathbf{r})\right|^2
\end{equation}
where $j$ is real. Currently we use $j\in\{0,1,2\}$, but other values of $j$ (including non-integers) are possible. However, $j<0$ and $j>2$ might cause numerical stability and normalization issues. Currently, $h(u; R)$ takes the form
\begin{equation}
    h(u; R) = \left(\frac{2}{\pi}\right)^{3/2} \frac{4}{4-\sqrt{2}}\frac{e^{-2u^2/R^2}}{R^3} \left(1 - e^{-2u^2/R^2}\right)
\end{equation}
Different functions could be chosen, but the important aspect of the smoothing functions is that convolving the density matrix with it yields a smoothed approximation to the spherically averaged density matrix at distance $R$, and that the level of smoothing increases as $R$ increases. These properties are necessary 1) to obey uniform scaling rules and 2) to ensure that $\rho^0(R; \mathbf{r})$ can be expanded efficiently using a Gaussian basis. To evaluate $H_j^0(\mathbf{r})$, the smoothed density matrix $\rho^0(R; \mathbf{r})$ is evaluated at a discrete set of distances $R_i$ and then interpolated over $R$ using a Gaussian basis; $H_j^0$ can then be evaluated analytically within the Gaussian basis, as described in the Supporting Information (Section S1). Note that under uniform scaling (eq~\ref{eq:uniform_scaling}), this feature scales as $\lambda^{3+j}$, i.e.
\begin{equation}
    H_j^0[n_\lambda](\mathbf{r})=\lambda^{3+j}H_j^0[n](\lambda\mathbf{r})
\end{equation}

The complexity of the SDMX features can be increased by also calculating expectation values of the gradients of the smoothed density matrix with respect to the length-scale $R$, as follows:
\begin{equation}
    H_j^{0\text{d}}(\mathbf{r}) = 4\pi \int \dd R\, R^{4-j} \left|\pdv{}{R} \rho^0(R; \mathbf{r})\right|^2
\end{equation}
In practice, these features can be obtained at almost no computational overhead compared to the original $H_j^0$ features because the bottleneck computational operations are the same as those needed to compute $H_j^0$ (see Supporting Information Section S1). This makes them a valuable way to increase the complexity of the featurization. The uniform scaling behavior of $H_j^{0\text{d}}$ is also $\lambda^{3+j}$.

The angular complexity of the SDMX features can be increased by computing (in addition to $\rho^0$) the following higher-order quantity
\begin{equation}
    \boldsymbol{\rho}^1(R; \mathbf{r}) = \int \dd[3]\mathbf{r}' \left[ \nabla h(|\mathbf{r}'-\mathbf{r}|; R) \right]  n_1(\mathbf{r}', \mathbf{r})
\end{equation}
which involves the gradients of $h$. Then, one can compute new features
\begin{equation}
    H_j^\text{1}(\mathbf{r}) = 4\pi \int \dd R\, R^{4-j} \left|\boldsymbol{\rho}^1(R; \mathbf{r})\right|^2
\end{equation}
These are more costly to compute than $H_j^0$ or $H_j^\text{0d}$, but they provide significantly more information to the model because they contain $\ell=1$ angular information. The uniform scaling behavior of $H_j^1$ is also $\lambda^{3+j}$.

Finally, one can combine the radial and Cartesian derivatives to construct one more type of feature
\begin{equation}
    H_j^\text{1d}(\mathbf{r}) = 4\pi \int \dd R\, R^{6-j} \left|\pdv{}{R} \boldsymbol{\rho}^1(R; \mathbf{r})\right|^2
\end{equation}
which also scales as $\lambda^{3+j}$ under uniform scaling and comes at very little additional computational cost if $\boldsymbol{\rho}^1(R; \mathbf{r})$ was already computed to evaluate $H_j^\text{1}(\mathbf{r})$.

We combine all of these features with semilocal features to learn the exchange functional, as described in Section~\ref{sec:methods}. As with all our previous functionals, these models obey the uniform scaling rule of the exact exchange energy as well as the uniform electron gas limit.

\section{Methods\label{sec:methods}}

\subsection{Model Details}

The exchange energy was fit using a Gaussian process as described in Section~\ref{sec:theory_gp}. One key difference from our previous work~\cite{bystrom_nonlocal_2023} is that instead of using an additive Gaussian process kernel and then mapping it to cubic splines after training~\cite{bystrom_nonlocal_2023}, we instead used a standard squared exponential kernel, which improves the flexibility of the model as explained in Appendix~\ref{app:flex_kernel}. This kernel has the form
\begin{equation}
    k(\mathbf{x},\mathbf{x'}) = \Sigma_0 C^2 n^{4/3} \left(n'\right)^{4/3}  \exp\left[-\frac{1}{2}\left(\mathbf{\bar{x}}-\mathbf{\bar{x}'}\right)^2\right] \label{eq:k_details}
\end{equation}
where $\bar{x}_i=x_i/l_i$ is the feature vector scaled by a set of length-scale hyperparameters $l_i$ and $\Sigma_0=2$ is a covariance hyperparameter. In eq~\ref{eq:k_details}, $n$ and $n'$ are the values of the density for the two grid points at which the features $\mathbf{x}$ and $\mathbf{x}'$ are evaluated, respectively, and $C=-\frac{3}{4}\left(\frac{3}{\pi}\right)^{1/3}$ is the uniform electron gas exchange prefactor~\cite{dirac_note_1930}. As explained in Appendix~\ref{app:ex_kernel_form}, the use of eq~\ref{eq:k_details} as the kernel results in an exchange functional of the form
\begin{align}
    E_\text{x}&=\int \dd[3]\mathbf{r}\,e_\text{x}(\mathbf{r})\label{eq:mlx_integral}\\
    e_\text{x}(\mathbf{r})&=Cn(\mathbf{r})^{4/3}g(\mathbf{x}(\mathbf{r}))\label{eq:mlx_form}
\end{align}
for some function $g(\mathbf{x})$ of the feature vector. Each hyperparameter $l_i$ was set to the standard deviation of $x_i$ over a random sample of grid points in the training set systems. Because evaluating the Gaussian process is computationally expensive, we fit a small, dense, feed-forward neural network to reproduce the Gaussian process after training was complete. It would also be possible to directly train a neural network to the training data, but the current approach was found to be easier for the time being and yielded sufficiently accurate and numerically stable models.

The feature vector for the final model was the following:
\begin{equation}
    \mathbf{x}=\{s, \alpha, \tilde{H}_1^0, \tilde{H}_2^0, \tilde{H}_1^\text{0d}, \tilde{H}_2^\text{0d}, \tilde{H}_0^1, \tilde{H}_1^1, \tilde{H}_2^1, \tilde{H}_0^{1\text{d}}, \tilde{H}_2^{1\text{d}}\}\label{eq:feature_set}
\end{equation}
In the above equation, $s=\frac{|\nabla n|}{2\left(3\pi^2\right)^{1/3}n^{4/3}}$ is the reduced density gradient. Here $\alpha=\frac{\tau-\tau_W}{\tau_0}$ is the iso-orbital indicator~\cite{sun_density_2013}, with $\tau=\frac{1}{2}\sum_i f_i |\nabla\phi_i|^2$ being the kinetic energy density, $\tau_W=\frac{|\nabla n|^2}{8n}$  the one-electron kinetic energy density, and $\tau_0=\frac{3}{10}(3\pi^2)^{2/3}n^{5/3}$  the uniform electron gas kinetic energy density. In addition, $\tilde{H}_j(\mathbf{r})=\gamma_jH_j/n(\mathbf{r})^{1+j/3}$ is a normalized version of the SDMX feature described in Section~\ref{sec:theory_sdmx}, with $\gamma_j$ chosen such that the value of $\tilde{H}_j$ is 1 for the uniform electron gas. All of these features are invariant under uniform scaling, meaning that for $n_\lambda(\mathbf{r})$ as defined in eq~\ref{eq:uniform_scaling},
\begin{equation}
    \mathbf{x}[n_\lambda](\mathbf{r})=\mathbf{x}[n](\lambda\mathbf{r})
\end{equation}
Because of this property, the learned exchange functional of eqs~\ref{eq:mlx_integral}--\ref{eq:mlx_form} obeys the uniform scaling rule of eq~\ref{eq:exx_uscale}.

The particular choice of features in eq~\ref{eq:feature_set} was determined by computing the covariance of the features over a sample of grid points in the training set and picking some of them that had low covariance with each other. The number of features was chosen to balance efficiency and accuracy. This approach for selecting the features was heuristic and not systematic, and it could likely be improved by a more systematic screening of features. However, as the purpose of this work is to outline the key concepts and capabilities of the band gap fitting method rather than to exhaustively optimize a functional, we leave this systematic testing to future work.

The implementation of the new SDMX features for Gaussian-type orbital calculations is described in the Supporting Information (Section S1). We have implemented the approach for isolated systems (all-electron and pseudopotential formalisms) and periodic systems (pseudopotential formalism with uniform grids). In future work, we will implement support for plane-wave DFT and all-electron periodic calculations.

Throughout the rest of this work, we will refer to models trained with the above feature set as CIDER24X, and we will refer to the NL-MGGA-DTR model from our previous work as CIDER23X~\cite{bystrom_nonlocal_2023}.

\subsection{Training Data}~\label{sec:model_details}

The training set for our models consisted entirely of molecular data and contained the following subsets:
\begin{itemize}
    \item The uniform electron gas exchange energy~\cite{dirac_note_1930}, which is effectively incorporated as an ``exact constraint'' by giving it zero noise in the Gaussian process model. To find the uniform electron gas feature vector, we used a combination of analytical and numerical integration with Mathematica~\cite{inc_mathematica_nodate} and numpy~\cite{harris_array_2020} to compute the uniform electron gas limits of the SDMX features.
    \item The molecular part of the ``DTR'' training set of our previous work~\cite{bystrom_nonlocal_2023}, which is a subset of 225 relative molecular energies from the GMTKN55 database.
    \item The atomization energies of the W4-11 database, which is a subdatabase of GMTKN55.
    \item The HOMO, LUMO, and HOMO-LUMO gaps of the molecules and atoms in W4-11.
    \item The ionization potentials and electron affinities of the G21IP, G21EA, and 3d-SSIP30 databases. G21IP and G21EA are subdatabases of GMTKN55 and contain main-group atoms and small molecules. 3d-SSIP30~\cite{luo_density_2014} is a database of 20 spin-state excitations and 10 ionization potentials of $3d$ transition metal atoms.
    \item Equations~\ref{eq:quadraticity_constraint} and~\ref{eq:exx_quadraticity} for the ionization potentials and electron affinities of G21IP, G21EA, and 3d-SSIP30.
\end{itemize}
The incorporation of training data for periodic systems will be addressed in future work. We trained two versions of CIDER24X: One with the energy level training data (i.e. the latter three subsets above), which we denote CIDER24X-e for shorthand; and one without the energy level training data, which we denote CIDER24X-ne for shorthand. Comparing these models, which otherwise were trained with identical hyperparameters, feature sets, and training data, allowed us to investigate the effects of energy level fitting on the model predictions.

\section{Results and Discussion\label{sec:results}}

To highlight the performance of the new SDMX features, we compare CIDER24X (with and without energy levels in the training data) with CIDER23X, which is the NL-MGGA-DTR model from our previous work~\cite{bystrom_nonlocal_2023} that is fit only to the total energy data. In each case, we substitute the machined-learned exchange model in place of exact exchange in the PBE0~\cite{adamo_toward_1999} functional:
\begin{equation}
    E_\text{xc} = \frac{1}{4} E_\text{x}^\text{CIDER} + \frac{3}{4}E_\text{x}^\text{PBE} + E_\text{c}^\text{PBE}~\label{eq:cider0}
\end{equation}
where $E_\text{x}^\text{CIDER}$ is the CIDER exchange energy,  $E_\text{x}^\text{PBE}$ is the PBE~\cite{perdew_generalized_1996} exchange energy, and $E_\text{c}^\text{PBE}$ is the PBE correlation energy. 
Except for Section~\ref{sec:polarons}, in which we vary the fraction of CIDER exchange, all results for CIDER23X and CIDER24X refer to eq~\ref{eq:cider0}. Therefore, ``perfect'' performance for these models means predicting the same energy and single-particle energy levels as PBE0.

Before proceeding, we summarize the shorthand for each functional explored in Sections~\ref{sec:mol_energy} and~\ref{sec:mol_gaps}:
\begin{itemize}
    \item \textbf{PBE}: The PBE functional~\cite{perdew_generalized_1996} (a nonempirical GGA) with D4 dispersion corrections~\cite{caldeweyher_generally_2019}.
    \item \textbf{r$^2$SCAN}: The r$^2$SCAN functional~\cite{furness_accurate_2020} (a nonempirical meta-GGA) with D4 dispersion corrections.
    \item \textbf{PBE0}: The PBE0 functional~\cite{adamo_toward_1999} (a nonempirical hybrid functional) with D4 dispersion corrections.
    \item \textbf{CIDER23X}: The PBE0 surrogate defined by eq~\ref{eq:cider0}, with the NL-MGGA-DTR CIDER functional~\cite{bystrom_nonlocal_2023} used as the replacement for exact exchange. The PBE0 D4 dispersion corrections were used.
    \item \textbf{CIDER24X-ne}: The PBE0 surrogate defined by eq~\ref{eq:cider0}, with the CIDER functional from this work---trained \emph{without} explicit energy level reference data---used as the replacement for exact exchange. See Section~\ref{sec:model_details} for details. The PBE0 D4 dispersion corrections were used.
    \item \textbf{CIDER24X-e}: Same as CIDER24X-ne, except that the surrogate exchange functional was fit to energy level exact constraints and reference data in addition to relative energies of molecules. See Section~\ref{sec:model_details} for details. The PBE0 D4 dispersion corrections were used.
\end{itemize}
For computational details, see Appendix~\ref{sec:comp_details}.

\subsection{Molecular Energetic Data}\label{sec:mol_energy}

Due to their limited descriptive power, semilocal functionals suffer from a trade-off between the accurate description of band gaps and bond energies~\cite{kovacs_what_2022}. As shown by Kov\'acs \emph{et al.}\cite{kovacs_what_2022}, empirically fitting a GGA or meta-GGA to more accurately capture cohesive energies of solids worsens the prediction of band gaps and vice versa, even for data on which the model is fit. This suggests that this shortcoming of semilocal functionals is a ``model capacity'' issue, meaning that semilocal functionals cannot accurately capture both band gaps and bond energies regardless of how they are constructed.

\begin{figure}
    \centering
    \includegraphics[width=\textwidth]{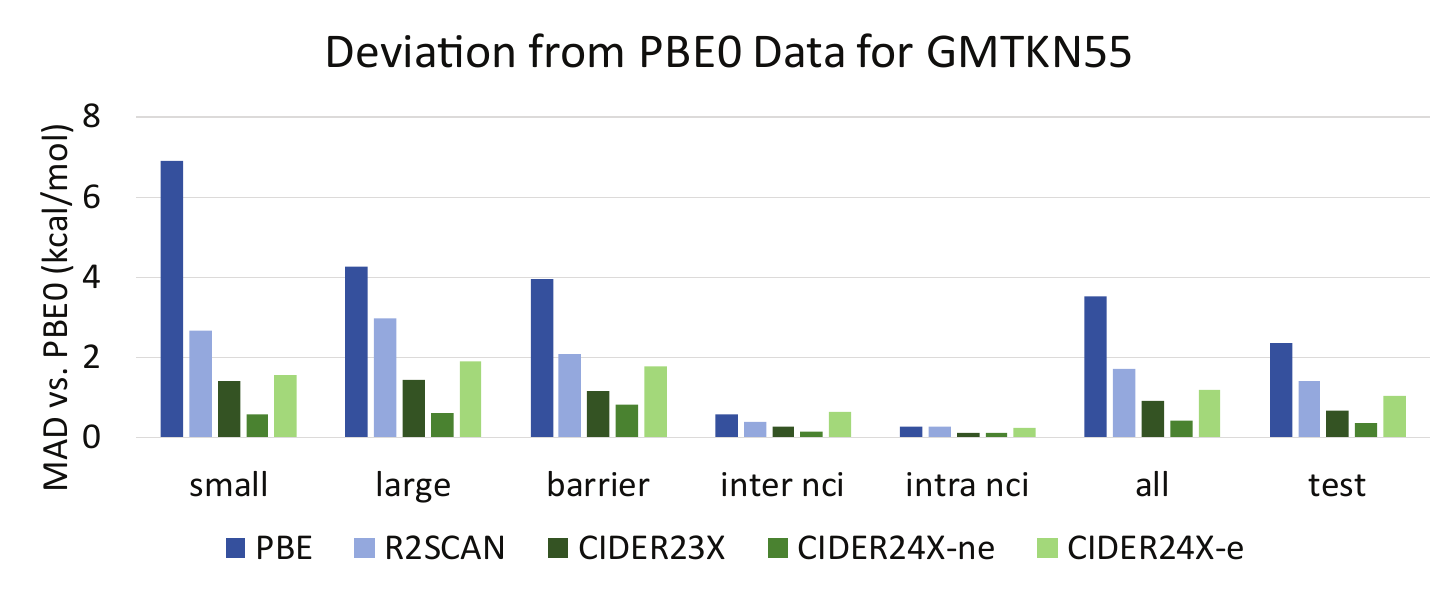}
    \caption{Mean absolute deviations (MAD) of different functionals from PBE0 (with D4 dispersion corrections) for each category in the GMTKN55 database, as well as the whole database (``all'') and the data not used for training CIDER24X (``test''). CIDER23X and CIDER24X refer to the PBE0/CIDER functional form of eq~\ref{eq:cider0}.}
    \label{fig:molecule_pbe0}
\end{figure}

\begin{figure}
    \centering
    \includegraphics[width=\textwidth]{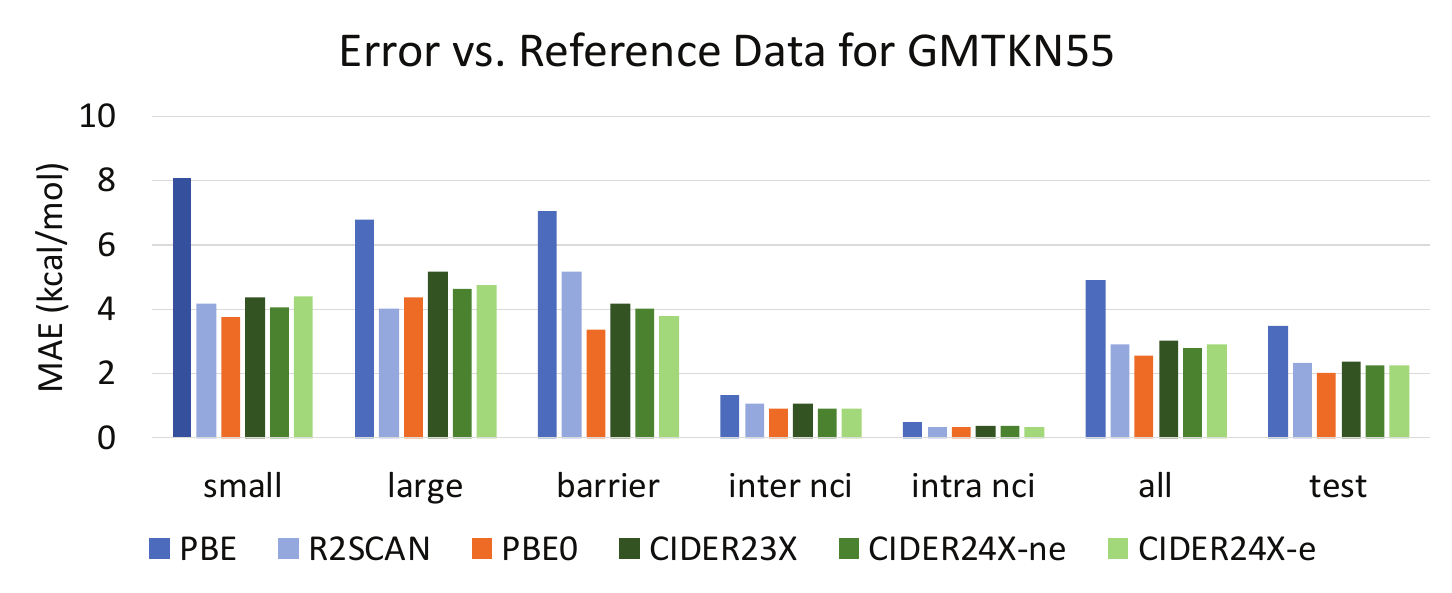}
    \caption{Mean absolute errors (MAE) of different functionals compared to experimental and quantum chemical reference values for each category of the GMTKN55 database, as well as the whole database (``all'') and the data not used for training CIDER24X (``test''). CIDER23X and CIDER24X refer to the PBE0/CIDER functional form of eq~\ref{eq:cider0}.}
    \label{fig:molecule_energies}
\end{figure}

Therefore, a key question for a new functional design approach is whether it can accurately predict energies and energy gaps using a single model. To this end, we benchmarked the performance of CIDER23X and CIDER24X on the GMTKN55 database~\cite{goerigk_look_2017}, a set of 1505 relative energies of molecules covering five categories of data: small-molecule reaction energies and other basis properties (``small''), large-molecule reaction energies and isomerization energies (``large''), barrier heights (``barrier''), intermolecular noncovalent interactions (``inter nci''), and intramolecular noncovalent interactions (``intra nci''). As shown in Figure~\ref{fig:molecule_pbe0}, the energy predictions made by CIDER24X without eigenvalue training (CIDER24X-ne) match PBE0 even more closely than CIDER23X, and in fact provide a chemically accurate match to PBE0 across the GMTKN55 database, with a mean absolute deviation (MAD) of 0.43 kcal/mol across all of GMTKN55. This suggests that the CIDER24X descriptors are more expressive and capable of fitting exact exchange than the previous features. When eigenvalue training data is incorporated (CIDER24X-e), the model agrees somewhat less closely with PBE0 than CIDER23X does. This suggests that even with these expressive features, there is a trade-off between the accuracy of energy predictions and eigenvalue predictions. However, the deviation from PBE0 is still fairly small on an absolute scale (1.2 kcal/mol across all of GMTKN55) and also typically smaller than the deviation between PBE0 and semilocal functionals like PBE and r$^2$SCAN. In addition, as shown in Figure~\ref{fig:molecule_energies}, CIDER23X and CIDER24X (with and without eigenvalue training data) all have very similar average deviations from the higher-level quantum chemistry and experimental reference values provided by GMTKN55. Thus, the worse match of CIDER24X-e to exact exchange does not translate into significantly worse property predictions relative to ``ground truth.'' Overall, even with eigenvalue training data, CIDER24X has comparable accuracy to the popular meta-GGA r$^2$SCAN for predicting relative energies of molecular systems.

Owing to the fact that CIDER24X-ne provides a significantly better match to PBE0 than CIDER24X-e across all five categories of GMTKN55, it seems likely that the error of CIDER24X-e is a model capacity issue, much like that seen with semilocal functionals. It is likely that this model capacity issue arises from the inability of the SDMX features to capture the long-range tail of the exchange operator, but further investigation is required to fully understand this issue. Regardless, as shown in the next section, CIDER24X-e provides a much better match to hybrid DFT HOMO-LUMO gaps than any semilocal functionals, so the accuracy trade-offs between energies and energies gaps is clearly less severe with CIDER24X.

\subsection{Molecular HOMO-LUMO Gaps}\label{sec:mol_gaps}

Having seen that we can fit an exchange functional to eigenvalue data while still obtaining reasonable predictions of molecular energies, we now show that the eigenvalue training actually results in improved band gaps. As shown in Table~\ref{tab:small_mol_gaps}, for the case of a few common small molecules, CIDER23X predicts HOMO-LUMO gaps that are similar to those of r$^2$SCAN and much smaller than those of PBE0, showing that training to energies alone is insufficient to achieve hybrid accuracy for band gaps. CIDER24X-ne also significantly underestimates these gaps compared to PBE0, though they are slightly larger than those of CIDER23X. CIDER24X-e, on the other hand, predicts HOMO-LUMO gaps in excellent agreement with PBE0, with all gaps matching to within 0.3 eV (compared to roughly 1.5-3 eV deviations for CIDER23X and CIDER24X-ne). This demonstrates that explicitly incorporating the eigenvalue training data is necessary to obtain comparable energy gaps to hybrid DFT; such accurate predictions do not arise simply from introducing more expressive features like SDMX.

\begin{figure}
    \centering
    \includegraphics[width=\textwidth]{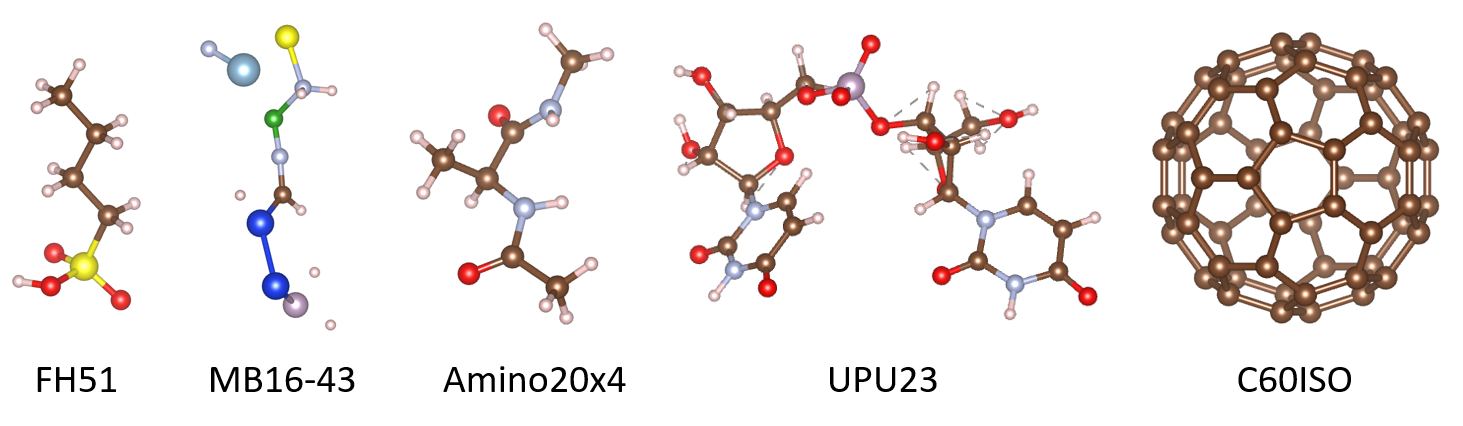}
    \caption{An example molecular system from each of the subdatabases used for HOMO-LUMO gap benchmarking. From left to right: FH51, small organic molecules; MB16-43, randomly generated main-group molecules; Amino20x4, amino acid conformers; UPU23, RNA-backbone conformers; C60ISO, \ce{C60} isomers. See Goerigk \emph{et al.}~\cite{goerigk_look_2017} for details on the subdatabases.}
    \label{fig:molecule_pictures}
\end{figure}

\begin{figure}
    \centering
    \includegraphics[width=\textwidth]{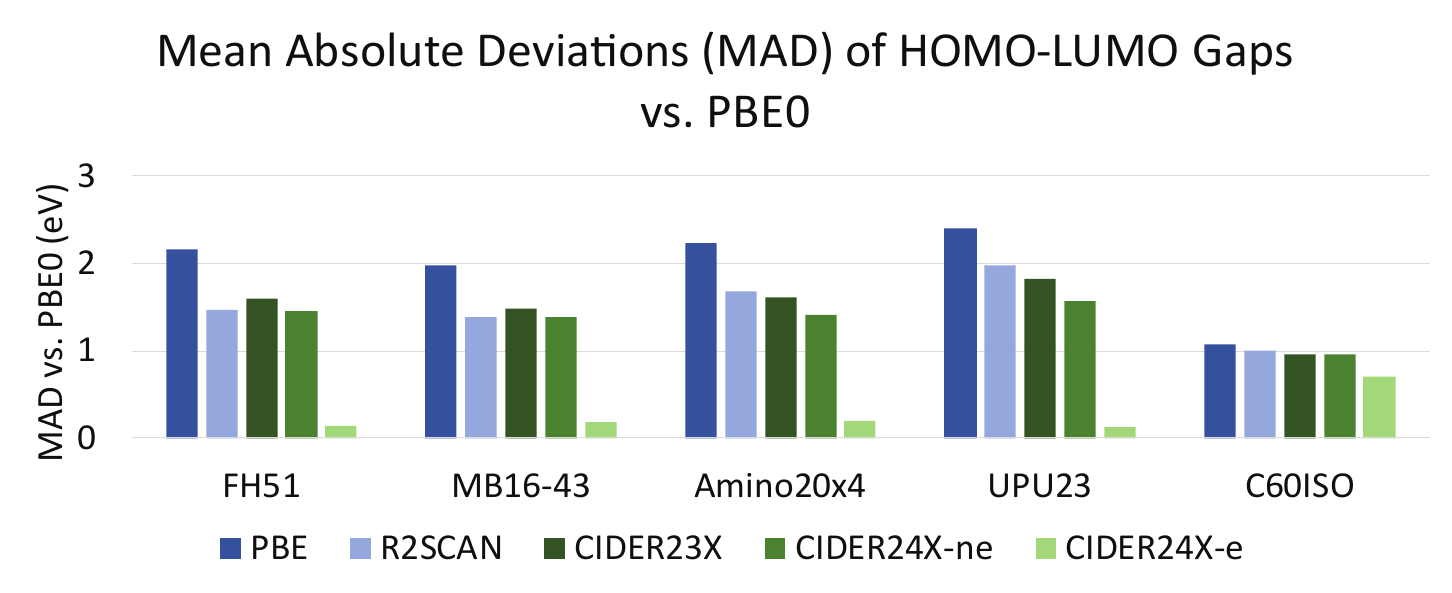}
    \caption{Mean absolute deviations (MAD) of different functionals vs. PBE0 for the prediction of HOMO-LUMO gaps for different classes of molecules from the GMTKN55 database~\cite{goerigk_look_2017}. Note that CIDER24X-e was not trained on the HOMO-LUMO gaps of these systems.}
    \label{fig:test_gaps}
\end{figure}

The energy levels of the simple small molecules above were a subset of the training data for CIDER24X-e, but as shown in Figure~\ref{fig:test_gaps}, the accurate HOMO-LUMO gap predictions are transferable to larger and more complex molecules whose eigenvalue data was not used to fit CIDER24X-e. (Some of these systems were still used for \emph{energy} training for CIDER24X-ne and CIDER24X-e, but not eigenvalue training.) These more complex molecules were taken from subdatabases of GMTKN55 and are illustrated in Figure~\ref{fig:molecule_pictures}. For every subdatabase except C60ISO (\ce{C60} isomers), CIDER24X-e has a mean absolute deviation of 0.2 eV or lower from PBE0. Even for C60ISO, CIDER24X-e exhibits better agreement with PBE0 than CIDER23X and semilocal DFT. We expect that the difficulty in describing the HOMO-LUMO gaps of \ce{C60} isomers arises from the delocalized nature of the aromatic bonds in these systems. In practice, we expect that long-range exchange contributions to the band gap will be negated by screening correlation effects in systems like these, which means they will be easier to handle with a full XC functional. This intuition is matched by the fact that PBE0 significantly overestimates the HOMO-LUMO gap of the \ce{C60} molecule compared to experiment. The PBE0 gap of \ce{C60} is 2.94 eV (as computed for this work with a def2-QZVPPD basis), while the experimental gap is 1.86 eV (via temperature-dependent microwave conductivity~\cite{rabenau_energy_1993}), suggesting that too much long-range exchange interaction is physically unrealistic in this type of system.

Overall, there is a clear advantage to explicitly fitting energy gaps and using the more powerful SDMX feature set introduced in this work. For molecular energetic data, the CIDER24X-e model has similar accuracy to r$^2$SCAN and only slightly lower accuracy than PBE0 (Figure~\ref{fig:molecule_energies}), while at the same time predicting energy gaps in excellent agreement with PBE0, especially compared to semilocal functionals (Figure~\ref{fig:test_gaps}). These results were obtained without computing the exact exchange energy, which is a promising indication that machine learning can help solve the band gap problem without the significant computational cost associated with hybrid DFT.

\begin{table}[]
    \centering
    \begin{tabular}{crrrrrr}
        System & PBE & r$^2$SCAN & PBE0 & CIDER23X & \makecell{CIDER24X-ne} & \makecell{CIDER24X-e} \\
        \hline
        \ce{CO} &	7.02 &	7.78 &	9.97 &	7.75 & 7.99 &	10.06 \\
        \ce{F2} &	3.63 &	4.52 &	7.79 &	4.71 & 5.26 &	7.77 \\
        \ce{HF} &	9.08 &	10.46 &	12.12 &	9.94 & 10.29 &	12.19 \\
        \ce{N2} &	8.29 &	9.06 &	11.57 &	9.12 & 9.60 &	11.86 \\
        \ce{H2O} &	6.56 &	7.74 &	9.12 &	7.29 & 7.57 &	9.23 \\
    \end{tabular}
    \caption{HOMO-LUMO gaps of small molecules (in eV). CIDER23X and CIDER24X denote substituting the respective exchange functionals into PBE0 in place of exact exchange. Note that the training data of CIDER24X-e contained (among other data) the HOMO-LUMO gaps of these systems.}
    \label{tab:small_mol_gaps}
\end{table}

\subsection{Polaron Formation Energies}~\label{sec:polarons}

\begin{table}[tbh]
    \centering
    \begin{tabular}{cccc}
        System & Exact Exchange & CIDER24X-ne Exchange & CIDER24X-e Exchange \\
        \hline
        \ce{MgO} Hole & 0.34 & 0.86 & 0.48 \\
        \ce{BiVO4} Electron & 0.14 & 0.48 & 0.19
    \end{tabular}
    \caption{Fraction of exact exchange and CIDER exchange necessary to enforce the piecewise linearity of the total energy (eq~\ref{eq:pwl_polaron}) for the hole polaron in MgO and the electron polaron in \ce{BiVO4}. The fractions of exact exchange are taken from Ref.~\cite{falletta_polarons_2022}.}
    \label{tab:exx_frac}
\end{table}

One of the most important applications of the eigenvalue-fitting technique in this work would be the prediction of electronic properties of solids. Because we have not yet implemented the use of solid-state training data, we do not expect the current model to consistently predict accurate solid-state band gaps. However, due to the excellent description of localized electron levels described in the previous section, we decided to test the ability of CIDER24X to describe the formation energies of polarons in solids, which are quasiparticles consisting of an electron (or hole) and the lattice distortion induced by the excess charge \cite{franchini2021NRM}. The formation energy of a polaron is the energy difference between the bulk solid with a delocalized charge in the valence band (hole polaron) or conduction band (electron polaron) and the distorted structure with the excess charge localized on the distortion. The accurate description of polarons with DFT is difficult due to self-interaction error \cite{falletta_many-body_2022,falletta_polarons_2022}. A semilocal functional like PBE will result in the polaron having positive formation energy: Rather than inducing a lattice distortion and localizing on the distorted site, the excess charge will simply delocalize, resulting in no polaron formation.

Models that mitigate self-interaction, such as DFT+$U$ and hybrid DFT, can be used to localize polarons, but the ideal parameterization of the model (e.g. the value of $U$ in DFT+$U$, the fraction of exact exchange in hybrid DFT) is system-dependent and needs to be determined. Falletta and Pasquarello introduced an approach to model polarons based on selecting the parameter of the functional to enforce the piecewise linearity condition of the total energy with respect to polaron occupation.\cite{falletta_many-body_2022,falletta_polarons_2022} Specifically, through Janak's theorem~\cite{janak_proof_1978}, the piecewise linearity of the total energy can be enforced by imposing the polaron level to be constant upon fractional electron occupation, namely
\begin{equation}
    \epsilon_\text{polaron}(q)=\epsilon_\text{polaron}(0)\label{eq:pwl_polaron}
\end{equation}
where $\epsilon_\text{polaron}(q)$ is the polaron level at a system net charge of $q$, calculated at the equilibrium polaron geometry with $q=+1$ for hole polarons or $q=-1$ for electron polarons. The above condition is non-empirical because it is a requirement for the piecewise linearity of the total energy with respect to electron number, which is a known behavior of the exact DFT energy~\cite{cohen_insights_2008}. This in turn yields ground state and transport properties of polarons that remain robust irrespective of the adopted functional.\cite{falletta_many-body_2022,falletta_polarons_2022,falletta_hubbard_2022,falletta2023PRB,falletta2024JAP} 

The CIDER24X functional can be used to localize polarons similarly to the way the PBE0($\alpha$) hybrid functional is used. In particular, the PBE0($\alpha$) functional is defined as 
\begin{equation}
    E_\text{xc}^{\text{PBE0}}(\alpha) = \alpha E_\text{x}^\text{exact} + (1-\alpha)E_\text{x}^\text{PBE} + E_\text{c}^\text{PBE}~\label{eq:pbe0_alpha}
\end{equation}
where $\alpha$ is the fraction of exact exchange energy $E_\text{x}^\text{exact}$ admixed to the semilocal exchange. As $\alpha$ increases, the electron polaron level $\epsilon_\text{polaron}(-1)$ becomes lower in energy relative to the delocalized conduction band edge. Similarly, the hole polaron level $\epsilon_\text{polaron}(+1)$ stabilizes with respect to the delocalized valence band edge. Generally, a dependence with $\alpha$ is also found for the neutral polaron level $\epsilon_\text{polaron}(0)$. Since the dependence of such polaron levels is generally roughly linear with $\alpha$~\cite{falletta_polarons_2022}, it is straightforward to identify the fraction of exact exchange such that eq~\ref{eq:pwl_polaron} is obeyed. Likewise, in the case of the CIDER functional, we choose the mixing fraction of CIDER24X-e such that eq~\ref{eq:pwl_polaron} is obeyed. We call this functional CIDER($\alpha$) for short:
\begin{equation}
    E_\text{xc}(\alpha) = \alpha E_\text{x}^\text{CIDER24X} + (1-\alpha)E_\text{x}^\text{PBE} + E_\text{c}^\text{PBE}~\label{eq:cider_alpha}
\end{equation}
We test this equation with both CIDER24X-ne and CIDER24X-e.

In this work, we focus on the hole polaron in \ce{MgO} and the electron polaron in \ce{BiVO4}. After determining the ideal functional parameterization via eq~\ref{eq:pwl_polaron}, there are two ways to determine the polaron formation energy. The first is to compute the formation energy from the polaron eigenvalue via~\cite{falletta_polarons_2022}
\begin{equation}
    E_\text{f}^\text{eigval}(q) = q(\epsilon_b-\epsilon_\text{polaron})+(E_\text{P}(0)-E_\text{B}(0))\label{eq:ef_eigval}
\end{equation}
where $q$ is the polaron charge and $E_\text{P}(q)$ and $E_\text{B}(q)$ are the total energies of the distorted and bulk crystal structures at charge $q$, respectively. The distorted structure is the equilibrium geometry obtained with the excess polaron charge $q$. The band edge $\epsilon_\text{b}$ is the valence band maximum for hole polarons and conduction band minimum for electron polarons. One can also use the traditional ``$\Delta$SCF'' approach~\cite{falletta_polarons_2022}
\begin{equation}
    E_\text{f}^{\Delta\text{SCF}}(q) = E_\text{P}(q) - E_\text{B}(0) + q\epsilon_b \label{eq:ef_dscf}
\end{equation}
For both approaches, the energies and eigenvalues must have finite-size corrections applied. We employ the correction scheme used by Falletta and Pasquarello~\cite{falletta_finite-size_2020}.
For a perfectly piecewise-linear functional, $E_\text{f}^\text{eigval}=E_\text{f}^{\Delta\text{SCF}}$, but the two values differ in general because eq~\ref{eq:pwl_polaron} is only one condition of piecewise linearity and does not enforce perfect piecewise linearity. Therefore, it is interesting to compare the two approaches. For further computational details, see Appendix~\ref{sec:comp_details}, and see Falletta and Pasquarello~\cite{falletta_polarons_2022} for details on the method for evaluating polaron formation energies.

Table \ref{tab:exx_frac} shows the fraction of exact and CIDER24X-e exchange necessary to fulfill the piecewise linearity condition of eq~\ref{eq:pwl_polaron}. The fraction of CIDER24X-e necessary to enforce this condition is somewhat larger than that of exact exchange, but not unreasonably so. The need for a 30-40\% larger mixing fraction suggests that CIDER is still capturing roughly two-thirds of the derivative discontinuity introduced by exact exchange. On the other hand, for CIDER24X-ne, which contains no eigenvalue training data, roughly three times as much CIDER exchange is required compared to exact exchange to enforce eq~\ref{eq:pwl_polaron}. This indicates that the physically realistic description of the polaron eigenvalues requires explicit eigenvalue fitting.

\begin{figure}[tbh]
    \centering
    \includegraphics[width=0.23\textwidth]{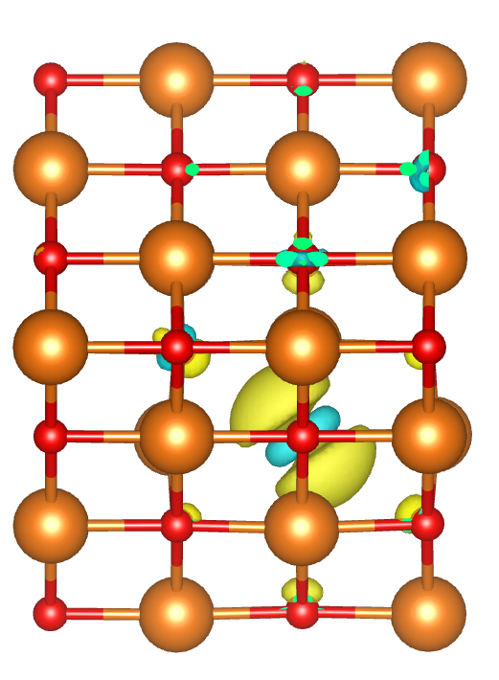}
    \includegraphics[width=0.27\textwidth]{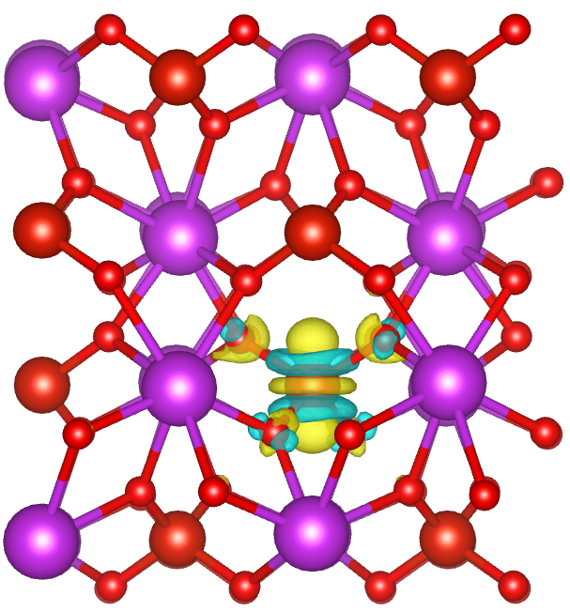}
    \caption{Charge distributions of the hole polaron in \ce{MgO} (left) and of the electron polaron in \ce{BiVO4} (right), obtained using the eq~\ref{eq:cider_alpha} with CIDER24X-e. Visualization performed using Vesta~\cite{Momma2008}. The CIDER24X-ne polaron charges are similarly localized, as shown in Supporting Information Figure S3.}
    \label{fig:polaron_chg_loc}
\end{figure}

\begin{table}[tbh]
    \centering
    \begin{subtable}[h]{\textwidth}
    \caption{Formation energies from eigenvalue method of eq~\ref{eq:ef_eigval}.}
    \begin{tabular}{ccccccc}
        System & $\gamma$DFT$^a$ & $\mu$DFT$^b$ & DFT+$U$$^c$ & PBE0($\alpha$)$^{a,d}$ & CIDER24X-ne($\alpha$)$^e$ & CIDER24X-e($\alpha$)$^e$ \\
        \hline
        \ce{MgO} & $-0.50$ & $-0.47$ & $-0.64$ & $-0.53$ & $-0.53$ & $-0.29$ \\
        \ce{BiVO4} & $-0.44$ & $-0.44$ & $-0.49$ & $-0.63$ & $-0.84$ & $-0.39$
    \end{tabular}
    \label{tab:polaron_formen_eigval}
    \end{subtable}
    \hfill\\
    \begin{subtable}[h]{\textwidth}
    \caption{Formation energies from $\Delta$SCF method of eq~\ref{eq:ef_dscf}.}
    \begin{tabular}{ccccccc}
        System & $\gamma$DFT$^a$ & $\mu$DFT$^b$ & DFT+$U$$^c$ & PBE0($\alpha$)$^{a,d}$ & CIDER24X-ne($\alpha$)$^e$ & CIDER24X-e($\alpha$)$^e$ \\
        \hline
        \ce{MgO} & $-0.42$ & $-0.67$ & $-0.57$ & $-0.50$ & $-0.54$ & $-0.23$ \\
        \ce{BiVO4} & $-0.59$ & $-0.62$ & $-0.73$ & $-0.71$ & $-1.38$ & $-0.72$
    \end{tabular}
    \label{tab:polaron_formen_dscf}
    \end{subtable}
    \caption{Formation energies (in eV) of the \ce{MgO} hole polaron and \ce{BiVO4} electron polaron computed with different piecewise-linear functionals. The PBE0 and CIDER functionals use the exchange mixing fraction of Table~\ref{tab:exx_frac}, and the other methods are parameterized as described in the table footnotes.}
    \begin{flushleft}
    $^a$Using the method and $\gamma$ parameter of Falletta and Pasquarello~\cite{falletta_polarons_2022}. The eigenvalue formation energy is taken from that work, and the $\Delta$SCF formation energy is from this work.\\
    $^b$Using the method and $U$ parameter of Falletta and Pasquarello~\cite{falletta2024JAP}. The eigenvalue formation energy is taken from that work, and the $\Delta$SCF formation energy is from this work.\\
    $^c$Using the method and $\mu$ parameter of Falletta and Pasquarello~\cite{falletta_hubbard_2022}. The eigenvalue formation energy is taken from that work, and the $\Delta$SCF formation energy is from this work.\\
    $^d$This functional corresponds to eq~\ref{eq:pbe0_alpha}.\\
    $^e$This functional corresponds to eq~\ref{eq:cider_alpha}.\\
    \end{flushleft}
    \label{tab:polaron_formen}
\end{table}

The key question is now whether enforcing eq~\ref{eq:pwl_polaron} causes the polaron charge to localize on the lattice distortion with a reasonable formation energy. Figure~\ref{fig:polaron_chg_loc} illustrates that this is indeed the case. For both the \ce{MgO} and \ce{BiVO4} polaron, the charge density difference between the charged and neutral distorted systems is highly localized on the site of the distortion (Figure~\ref{fig:polaron_chg_loc}). Figure~\ref{fig:polaron_chg_loc} shows the charge distribution for CIDER24X-e, but the results for CIDER24X-ne are nearly identical and shown in Supporting Information Figure S3. Next, we compare the polaron formation energies obtained with CIDER with those obtained with piecewise-linear functionals, including PBE0($\alpha$), $\gamma$DFT \cite{falletta_many-body_2022,falletta_polarons_2022}, $\mu$DFT \cite{falletta2024JAP}, and DFT+$U$. Specifically, the $\gamma$DFT and $\mu$DFT functionals are based on adding a weak local potential to the semilocal PBE Hamiltonian to favor polaron localization. The parameters of these functionals, $\gamma$ and $\mu$, respectively, are fixed in order to ensure the piecewise-linearity condition in eq~\ref{eq:pwl_polaron}. Similarly, in DFT+$U$ calculations, the parameter $U$ is applied to the orbitals constituting the polaron state to enforce eq~\ref{eq:pwl_polaron}.

Table~\ref{tab:polaron_formen} shows the polaron formation energies computed with a variety of different self-interaction correction methods including hybrid DFT and CIDER24X. As shown in Table~\ref{tab:polaron_formen_eigval}, the CIDER24X-ne formation energies, as computed by eq~\ref{eq:ef_eigval}, are in reasonable agreement with PBE0 (with deviations of 0.00 and 0.21 eV for \ce{MgO} and \ce{BiVO4}, respectively) in spite of the unphysically large fractions of CIDER exchange required to enforce piecewise linearity (Table~\ref{tab:exx_frac}). The CIDER24X-e formation energies are also negative, indicating a stable polaron, though CIDER24X-e underestimates both polaron formation energies by 0.24 eV compared to PBE0.

In principle, the $\Delta$SCF formation energies in Table~\ref{tab:polaron_formen_dscf} should be the same as those in Table~\ref{tab:polaron_formen_eigval}; however, this is not the case due to the imperfect piecewise linearity of the functionals. Previously developed approaches ($\gamma$DFT, $\mu$DFT, DFT+$U$, and PBE0) have deviations between eqs~\ref{eq:ef_eigval} and~\ref{eq:ef_dscf} ranging up to 0.24 eV for these polarons, though the larger of the two PBE0 deviations is only 0.08 eV (for \ce{BiVO4}). This problem is much more extreme for CIDER24X-ne, which does not have any training data for energy levels. For \ce{BiVO4}, there is a deviation between eqs~\ref{eq:ef_eigval} and~\ref{eq:ef_dscf} of 0.54 eV, with the CIDER24X-ne formation energy disagreeing with PBE0 by 0.67 eV. The problem is less severe for CIDER24X-e, which is fit to energy level data and trained to be roughly piecewise quadratic in orbital occupation. For \ce{BiVO4}, eqs~\ref{eq:ef_eigval} and~\ref{eq:ef_dscf} deviate by 0.33 eV (still significant but closer to the deviations for other methods). In addition, with eq~\ref{eq:ef_dscf}, CIDER24X-e deviates by only 0.01 eV from PBE0 for \ce{BiVO4}. This hints that the energy level training, while still in need of further development, might give CIDER24X-e more physically realistic behavior for fractional charges. However, since we have only studied two polarons, further work is required to say this with certainty.

In general, more research is required to understand the behavior of this new class of functionals for polaron and defect problems. In particular, we expect that fitting our exchange model to both energy data and band gaps of solids will significantly improve the agreement between CIDER24X-e and PBE0. As it stands, however, CIDER24X-e already provides a reasonably physically realistic description of these polarons, which is an initial indication of the transferability of the CIDER model across chemical space and its ability to tackle problems beyond simple benchmark datasets.

\section{Conclusion\label{sec:conclusion}}

We have introduced a general framework for fitting XC functionals to both total energy and single-particle energy levels using machine learning. This approach leverages Janak's theorem to connect the DFT eigenvalues to the derivative of the energy with respect to orbital occupation. We have also introduced a new class of features of the density matrix, called SDMX features, that can be used to learn the XC functional. The SDMX features use nonlocal convolutions of the density matrix to obtain approximate characterizations of the exchange hole without generating the full exact exchange operator required by hybrid DFT.

To illustrate the capabilities of this new approach and feature set, we trained a nonlocal exchange functional to molecular energies and single-particle energy levels and demonstrated that it is possible to retain the accuracy of previous models for energy data while drastically improving the accuracy of energy gaps. Our CIDER24X-e model is as accurate as r$^2$SCAN for relative energies in the GMTKN55 database, but it predicts larger HOMO-LUMO gaps in good agreement with hybrid DFT. For simple small molecules as well as more complex systems like RNA backbone chains, CIDER24X-e gaps agree with the reference PBE0 gaps with a mean absolute deviation of 0.2 eV or less. Despite being trained solely on molecular data, CIDER24X-e is sufficiently chemically transferable to give a physically realistic description of polarons in solids, such as the hole polaron in \ce{MgO} and the electron polaron in \ce{BiVO4}.

While the framework presented here was illustrated using only the exchange functional and a Gaussian process model, the method is extensible to learning the full XC functional as well as to other machine learning approaches like neural networks. A full XC functional trained using this approach could provide a route toward accurately describing the electronic properties of heterogeneous systems like solid surfaces and interfaces. Combined with careful code optimization for the feature generation, it could even be feasible to attain such accuracy at computational cost similar to semilocal DFT, since our models do not require evaluating the computationally expensive exact exchange energy. This work lays the foundation for fitting full XC functionals to single-particle energy levels.

\appendix

\section{Derivation of Piecewise Quadraticity Constraint on the Exchange Functional}\label{app:quad}

The exact exchange functional, as given by eq~\ref{eq:exact_exchange_energy}, can be written in terms of the orbitals and their occupation numbers as
\begin{equation}
    E_\text{x}^\text{exact} = -\frac{1}{2} \sum_\sigma \sum_i \sum_j f_{\sigma,i} f_{\sigma,j} \int \dd[3]\mathbf{r} \int \dd[3]\mathbf{r}' \frac{\phi_{\sigma,i}^*(\mathbf{r}) \phi_{\sigma,j}(\mathbf{r}) \phi_{\sigma,i}(\mathbf{r}') \phi_{\sigma,j}^*(\mathbf{r}')}{|\mathbf{r}-\mathbf{r}'|}
\end{equation}
Because $E_\text{x}^\text{exact}$ is spin-separable, i.e. \begin{equation}
    E_\text{x}[\{\phi_{\uparrow,i}\},\{\phi_{\downarrow,i}\}]=\sum_{\sigma=\uparrow,\downarrow} E_\text{x}[\{\phi_{\sigma,i}\}]
\end{equation}
we can focus on a single spin-channel without loss of generality:
\begin{equation}
    E_\text{x}^\text{exact} = -\frac{1}{2} \sum_i \sum_j f_{i} f_{j} \int \dd[3]\mathbf{r} \int \dd[3]\mathbf{r}' \frac{\phi_i^*(\mathbf{r}) \phi_j(\mathbf{r}) \phi_i(\mathbf{r}') \phi_j^*(\mathbf{r}')}{|\mathbf{r}-\mathbf{r}'|}\label{eq:exx_single_channel}
\end{equation}
We wish to investigate the behavior of this quantity as the number of electrons in the system $N$ changes, and specifically the behavior for fractional electrons ($N\notin\mathbb{Z}$). For simplicity, we denote the initial electron number as $N=0$ (though the derivation applies for any initial number of electrons) and explore the effect of increasing the electron count to $N=1$. We will assume that
\begin{equation}
    \pdv{\phi_i(\mathbf{r})}{N}=0\,\,\,\,\,\,N\in(0, 1)
\end{equation}
This is equivalent to the frozen orbital approximation employed in the derivation of Koopmans' theorem. Under this assumption, all changes to the density arise from changes to the orbital occupation numbers, which obey the following rules for $N\in[0,1]$ (assuming non-degeneracy of the orbitals):
\begin{equation}
    f_{i}(N)=\begin{cases}
    1 & \epsilon_{i}<\epsilon_\text{CB}\\
    N & \epsilon_{i}=\epsilon_\text{CB}\\
    0 & \epsilon_{i}>\epsilon_\text{CB}
    \end{cases}
\end{equation}
Under these conditions, eq~\ref{eq:exx_single_channel} becomes (for $N\in[0,1]$)
\begin{align}
    E_\text{x}^\text{exact} =& -\frac{1}{2} \sum_{i:\epsilon_i<\epsilon_\text{CB}} \sum_{j:\epsilon_j<\epsilon_\text{CB}} \int \dd[3]\mathbf{r} \int \dd[3]\mathbf{r}' \frac{\phi_i^*(\mathbf{r}) \phi_j(\mathbf{r}) \phi_i(\mathbf{r}') \phi_j^*(\mathbf{r}')}{|\mathbf{r}-\mathbf{r}'|}\notag\\
    &-N\sum_{i:\epsilon_i<\epsilon_\text{CB}}\int \dd[3]\mathbf{r}\int \dd[3]\mathbf{r}' \frac{\phi_i^*(\mathbf{r}) \phi_\text{CB}(\mathbf{r}) \phi_i(\mathbf{r}') \phi_\text{CB}^*(\mathbf{r}')}{|\mathbf{r}-\mathbf{r}'|}\notag\\
    &-\frac{N^2}{2}\int \dd[3]\mathbf{r}\int \dd[3]\mathbf{r}' \frac{\phi_\text{CB}^*(\mathbf{r}) \phi_\text{CB}(\mathbf{r}) \phi_\text{CB}(\mathbf{r}') \phi_\text{CB}^*(\mathbf{r}')}{|\mathbf{r}-\mathbf{r}'|}
\end{align}
which is quadratic in $N$. Thus, we have shown that the exact exchange energy is piecewise quadratic under the frozen orbital approximation, with the quadratic regions between consecutive integer electron numbers.

\section{Standard Versus Additive Squared-Exponential Gaussian Process Kernels}\label{app:flex_kernel}

One of the most popular and useful kernel functions for Gaussian processes (and the one employed in this work) is the squared exponential kernel, which for vector inputs takes the form
\begin{equation}
    k(\mathbf{x},\mathbf{x'}) = \Sigma_0 \exp\left[-\frac{1}{2}\left(\mathbf{\bar{x}}-\mathbf{\bar{x}'}\right)^2\right] \label{eq:k_app}
\end{equation}
where $\bar{x}_i=x_i/l_i$ is the feature vector scaled by a set of length-scale hyperparameters $l_i$, and $\Sigma_0$ is a scalar hyperparameter.

One of the drawbacks of Gaussian processes is that the computational cost of evaluating them scales linearly with the number of training points (or, in the case of the modified models in Section~\ref{sec:fit_energy}, the number of control points). This can be computationally costly, so it would be preferable to map the Gaussian process to a more efficient model once training is complete. One possible map is a cubic spline, which can be evaluated efficiently but becomes impractical to store and evaluate for higher-dimensional functions ($d>4$). In our previous work~\cite{bystrom_cider_2022,bystrom_nonlocal_2023}, to make this cubic spline map practical, we used an additive kernel~\cite{duvenaud_additive_2011} that is a linear combination of terms that each depend on a limited number of features. For example, a cubic additive kernel would be
\begin{equation}
    k_\text{additive}(\mathbf{x},\mathbf{x}')=\Sigma_0\sum_{i=1}^{N_\text{feat}}\sum_{j=i+1}^{N_\text{feat}}\sum_{m=j+1}^{N_\text{feat}}k_i(x_i,x_i')k_j(x_j,x_j')k_m(x_m,x_m')\label{eq:k_additive}
\end{equation}
with each scalar kernel function being
\begin{equation}
    k_i(x,x')=\exp[-\frac{1}{2}\left(\frac{x-x'}{l_i}\right)^2]\label{eq:basekernel}
\end{equation}
where $l_i$ are hyperparameters. Equation~\ref{eq:k_additive} yields a predictive function that can be mapped to a sum of three-dimensional cubic splines~\cite{bystrom_cider_2022}, and therefore can be evaluated efficiently. However, restricting the kernel in this way limits the expressiveness of the model, as only linear combinations of three-feature functions can be described. For example, if the feature vector is $\mathbf{X}=(a,b,c,d)$, eq~\ref{eq:k_additive} could successfully fit $f(\mathbf{x})=abc+bcd$ but not $f(\mathbf{x})=abcd$. Equation~\ref{eq:k_app}, on the other hand, can be used to fit both functions exactly.

In our previous models like CIDER23X, we found that the limitations imposed by eq~\ref{eq:k_additive} did not impede training accurate models in practice, and it was worthwhile to use the additive kernel so that the model could be mapped to cubic splines after training. In this work, however, we opted for the improved flexibility of eq~\ref{eq:k_app}, and instead of mapping to cubic splines, we mapped the final model to a neural network to improve computational efficiency. A more through benchmark of the trade-offs between the additive and standard kernels would be an interesting subject of future research, but it is not critical to the key points of this work.

\section{Exchange Kernel Functional Form}~\label{app:ex_kernel_form}

Equation~\ref{eq:k_details} is the kernel used in this work, and the predictive function for a given kernel is given by eq~\ref{eq:gp_sum_formula}. Substituting eq~\ref{eq:k_details} into eq~\ref{eq:gp_sum_formula} gives
\begin{align}
    f(\mathbf{x}_*) &= \sum_a k(\mathbf{x}_*, \mathbf{\tilde{x}}_a) \alpha_a \\
    &= \sum_a \left\{\Sigma_0 C^2 n_*^{4/3} \left(\tilde{n}_a\right)^{4/3}  \exp\left[-\frac{1}{2}\left(\mathbf{\bar{x}_*}-\mathbf{\bar{\tilde{x}}_a}\right)^2\right]\right\} \alpha_a \\
    &= C n_*^{4/3} \sum_a \exp\left[-\frac{1}{2}\left(\mathbf{\bar{x}_*}-\mathbf{\bar{\tilde{x}}_a}\right)^2\right] \left[ \Sigma_0 C (\tilde{n}_a)^{4/3} \alpha_a \right] \\
    &= C n_*^{4/3} g(\mathbf{x}_*)
\end{align}
In the above equations, $\alpha_a$ is given by eq~\ref{eq:gp_predictive}, $n_*$ and $\tilde{n}_a$ are the densities at the grid points where $\mathbf{x}_*$ and $\mathbf{\tilde{x}}_a$ are evaluated, respectively, and $\mathbf{\bar{x}}_*$ and $\mathbf{\bar{\tilde{x}}}_a$ are the feature vectors normalized by the length-scale hyperparameters, i.e. $\bar{x}_i=x_i/l_i$. In the final line, we see that $f(\mathbf{x}_*)$ is the product of the LDA exchange energy density~\cite{dirac_note_1930} and some learned function of the feature vector $g(\mathbf{x}_*)$. Since $f(\mathbf{x}_*)$ in our case is the exchange energy density, this confirms that eq~\ref{eq:mlx_form} holds.

\section{Computational Details}\label{sec:comp_details}

All DFT calculations were performed using PySCF~\cite{sun_pyscf_2018,sun_recent_2020}.  Benchmark calculations of GMTKN55~\cite{goerigk_look_2017,peverati_fitting_2021} molecules were performed with the def2-QZVPPD basis set~\cite{weigend_gaussian_2003,rappoport_property-optimized_2010} and def2-universal-jkfit auxiliary basis~\cite{weigend_hartreefock_2008} for the Coulomb energy. Hybrid DFT calculations were performed using the seminumerical exchange (SGX) module in PySCF. D4 dispersion corrections~\cite{caldeweyher_generally_2019} were used for all molecular DFT calculations, with PBE0 D4 parameters used for the CIDER functionals. Other calculation settings were identical to those in our previous work~\cite{bystrom_nonlocal_2023} (Appendix A2) unless otherwise stated.

For the MgO and BiVO4 polaron calculations, the GTH-PBE pseudopotentials~\cite{goedecker_separable_1996,hartwigsen_relativistic_1998,krack_pseudopotentials_2005} and DZVP-MOLOPT-SR-GTH basis sets~\cite{vandevondele_gaussian_2007} were used. For Mg, Bi, V, and O, the 2, 5, 13, and 6-electron pseudopotentials were used, respectively. For \ce{MgO}, a 64-atom supercell with a $81\times 81\times 81$ uniform grid was used for the calculation (both for constructing the density and integrating the XC energy and potential), and a $103\times 103\times 117$ uniform grid was used for 96-atom \ce{BiVO4} supercell calculations. The bulk and distorted structures were obtained from the work of Falletta and Pasquarello~\cite{falletta_many-body_2022,falletta_polarons_2022}, with the distorted structure relaxed using PBE0($\alpha$) with $\alpha$ chosen to satisfy eq~\ref{eq:pwl_polaron}. These structures were used without further optimization with the CIDER functionals. Only the $\Gamma$ point was sampled. The finite-size corrections described in Ref.~\cite{falletta_finite-size_2020} and used in Ref.~\cite{falletta_many-body_2022,falletta_polarons_2022} were used to correct for the effects of localized ionic polarization and excess charges in a periodic supercell model system. The formation energy of the polarons was then evaluated as described by Falletta and Pasquarello~\cite{falletta_many-body_2022,falletta_polarons_2022}.

All DFT calculations, isolated and periodic, including those using CIDER functionals, were performed self-consistently with static geometries.

\begin{acknowledgement}
This work was supported by the National Defense Science and Engineering Graduate (NDSEG) Fellowship Program under contract FA9550-21-F-0003, the Camille and Henry Dreyfus Foundation Grant No. ML-22-075, the Department of Navy award N00014-20-1-2418 issued by the Office of Naval Research, the US Department of Energy, Office of Basic Energy Sciences Award No.\ DE-SC0022199, and Robert Bosch LLC. Computational resources were provided by the Harvard University FAS Division of Science Research Computing Group.
\end{acknowledgement}

\begin{suppinfo}
The supporting information provides technical details for evaluating the SDMX features, plots of the polaron energy levels with respect to the fraction of CIDER24X-e exchange, and the charge distributions of the polarons computed with CIDER24X-ne.
\end{suppinfo}

\bibliography{references}

\end{singlespace} 
\end{document}


\preprint{APS/123-QED}

\title{Addressing the Band Gap Problem with a Machine-Learned Exchange Functional} 

\author{Kyle Bystrom}
\email{kylebystrom@g.harvard.edu}
\affiliation{Harvard John A. Paulson School of Engineering and Applied Sciences}
\author{Boris Kozinsky}
\email{bkoz@g.harvard.edu}
\affiliation{Harvard John A. Paulson School of Engineering and Applied Sciences}
\affiliation{Robert Bosch LLC Research and Technology Center, Cambridge, MA, USA}

\date{\today}


\onecolumngrid

\setcounter{page}{1}

\MakeTitle{Supplementary Information for ``Addressing the Band Gap Problem with a Machine-Learned Exchange Functional''}{Kyle Bystrom, Boris Kozinsky}

\beginsupplement

\section{Evaluation of SDMX Features}

The SDMX features are implemented within the PySCF program~\cite{sun_pyscf_2018,sun_recent_2020}, which uses Gaussian-type orbital basis sets. Each orbital is indexed by four numbers: $A$, the index of the atom on which the orbital is centered; $n$, the radial function index; and $\ell$ and $m$, the principal and azimuthal quantum numbers of the spherical harmonics. This combination of indices will be abbreviated by a single index $\mu=(A,n,\ell,m)$. We will also use a second atomic orbital index $\nu=(A',n',\ell',m')$. Each orbital takes the form $\chi_\mu(\mathbf{r})$, as given by the following equations:
\begin{align}
    \chi_\mu(\mathbf{r}) &= g_{An\ell}(|\mathbf{r}_A|) Y_{\ell m}(\hat{\mathbf{r}}_A) \label{eq:gto} \\
    g_{An\ell}(r) &= r^\ell \sum_j C_{An\ell,j} e^{-\alpha_j r^2}
\end{align}
with $\hat{\mathbf{r}}$ the direction of the vector $\mathbf{r}$, $Y_{\ell m}$ the real spherical harmonics, $\mathbf{R}_A$ the nuclear coordinate of atom $A$, and $\mathbf{r}_A=\mathbf{r}-\mathbf{R}_A$.

We will denote the Fourier transforms of these orbitals as
\begin{equation}
    \tilde{\chi}_\mu(\mathbf{k})=\mathcal{F}\{\chi_\mu(\mathbf{r})\}(\mathbf{k}) \label{eq:gto_fft}
\end{equation}
Because we are using Gaussian-type orbitals, eq~\ref{eq:gto_fft} also yields Gaussian-type orbitals in reciprocal space and can be evaluated analytically.

The molecular orbitals $\phi_i(\mathbf{r})$ are given in terms of the atomic basis functions as
\begin{equation}
    \phi_i(\mathbf{r}) = \sum_{\mu} c_{i\mu} \chi_\mu(\mathbf{r})
\end{equation}
The density matrix is then given by
\begin{align}
    n_1(\mathbf{r}, \mathbf{r}') &= \sum_i f_i \phi_i^*(\mathbf{r}) \phi_i(\mathbf{r}')\label{eq:dm_mo}\\
    n_1(\mathbf{r}, \mathbf{r}') &= \sum_{\mu\nu} P_{\mu\nu} \chi_\mu^*(\mathbf{r}) \chi_\nu(\mathbf{r}')\label{eq:dm_ao}
\end{align}
with the orbital-space density matrix $\mathbf{P}$ given by
\begin{equation}
    P_{\mu\nu} = \sum_i f_i c_{i\mu}^* c_{i\nu}
\end{equation}

There are two key steps to evaluating the SDMX features of Section 2.3. The first is to evaluate the smoothed density matrices $\rho^0(R; \mathbf{r})$ (for $H_j(\mathbf{r})$ and $H_j^\text{d}(\mathbf{r})$) or $\boldsymbol{\rho}^1(R; \mathbf{r})$ (for $H_j^1(\mathbf{r})$ or $H_j^{1\text{d}}(\mathbf{r})$) on a real-space quadrature grid $\mathbf{r}$ for a given set of distance parameters $R$. Plugging eqs~\ref{eq:dm_mo} and~\ref{eq:dm_ao} into eqs 50 and 55 from the main text, these quantities can be written in terms of the molecular orbitals or density matrix as
\begin{align}
    \rho^0(R; \mathbf{r}) &= \sum_{i} f_i \phi_i^*(\mathbf{r}) \int \dd[3]\mathbf{r}' h(|\mathbf{r}'-\mathbf{r}|; R) \phi_i(\mathbf{r}') \label{eq:rho0_mo} \\
    \rho^0(R; \mathbf{r}) &= \sum_{\mu\nu} P_{\mu\nu} \chi_\mu^*(\mathbf{r}) \int \dd[3]\mathbf{r}' h(|\mathbf{r}'-\mathbf{r}|; R) \chi_\nu(\mathbf{r}') \label{eq:rho0_ao} \\
    \boldsymbol{\rho}^1(R; \mathbf{r}) &= \sum_{i} f_i \phi_i^*(\mathbf{r}) \nabla\left( \int \dd[3]\mathbf{r}' h(|\mathbf{r}'-\mathbf{r}|; R) \phi_i(\mathbf{r}') \right) \label{eq:rho1_mo} \\
    \boldsymbol{\rho}^1(R; \mathbf{r}) &= \sum_{\mu\nu} P_{\mu\nu} \chi_\mu^*(\mathbf{r}) \nabla \left(\int \dd[3]\mathbf{r}' h(|\mathbf{r}'-\mathbf{r}|; R) \chi_\nu(\mathbf{r}')\right) \label{eq:rho1_ao}
\end{align}
The algorithm for evaluating eqs~\ref{eq:rho0_mo}--\ref{eq:rho1_ao} necessarily depends on the choice of basis set and periodicity of the system, so we discuss the different algorithms for isolated and periodic systems in Sections~\ref{sec:si_isolated} and~\ref{sec:si_periodic}, respectively.

The second step, given a way of evaluating the smoothed density matrices, is to perform the integrals of eqs 51, 54, 56, and 57 to obtain the SDMX features. To do so, we choose to evaluate the smoothed density matrices on a logarithmic grid for $R$
\begin{equation}
    R_j = R_0 \lambda^{j/2}, \,\,\,\,j=0,1,...,N_\text{dist}
\end{equation}
The parameters $\lambda$, $R_0$, and $N_\text{dist}$ are adjustable. In this work, we choose $\lambda=1.8$ and $R_0$ and $N_\text{dist}$ to make sure that the minimum and maximum $R_j$ cover the smallest and largest relevant length scales in the chemical system. Using calculated values of $\rho^0(R; \mathbf{r})$ and $\boldsymbol{\rho}^1(R; \mathbf{r})$ for each $R_j$, we then use interpolation to approximate the full functions $\rho^0(R; \mathbf{r})$ and $\boldsymbol{\rho}^1(R;, \mathbf{r})$ as a linear combination of Gaussians:
\begin{align}
    \rho^0(R; \mathbf{r}) &\approx \sum_{j=0}^{N_\text{dist}} D_{0j}(\mathbf{r}) e^{-\alpha_j R^2} \label{eq:interp_p0} \\
    \rho^1_\gamma(R; \mathbf{r}) &\approx \sum_{j=0}^{N_\text{dist}} D_{1\gamma j}(\mathbf{r}) e^{-\alpha_j R^2} \label{eq:interp_p1}
\end{align}
where $\alpha_j=2/R_j^2$ and $\gamma=x,\,y$, or $z$. The coefficients are determined by solving the systems of linear equations
\begin{align}
    \rho^0(R_i)&=\sum_j D_{0j} e^{-\alpha_j R_i^2} \\
    \rho^1_\gamma(R_i)&=\sum_j D_{1\gamma j} e^{-\alpha_j R_i^2}
\end{align}
Using the fact that Gaussians can be easily integrated analytically, the integrals 51, 54, 56, and 57 can then be evaluated from eqs~\ref{eq:interp_p0} and~\ref{eq:interp_p1} to obtain the SDMX features. Note that as $\lambda\rightarrow 1$, this approach becomes exact as the interpolation grid becomes infinitely dense (though ill-conditioning issues in the interpolation scheme can cause problems for small $\lambda$).

\subsection{Isolated Systems}\label{sec:si_isolated}

To evaluate eqs~\ref{eq:rho0_mo}--\ref{eq:rho1_ao}, one needs the convolutions of the atomic orbitals by every distance parameter $R$. We will denote these convolutions as
\begin{equation}
    \Gamma_\mu(\mathbf{r}; R) = \int \dd[3]\mathbf{r}' h(|\mathbf{r}'-\mathbf{r}|; R) \chi_\nu(\mathbf{r}')\label{eq:gamma_conv}
\end{equation}
We need $\Gamma_\mu(\mathbf{r}; R)$ to evaluate $\rho^0$ and its gradient to evaluate $\boldsymbol{\rho}^1$. Since $h(|\mathbf{r}'-\mathbf{r}|; R)$ is spherically symmetric, every convolution $\Gamma_\mu$ has the same angular dependence (spherical harmonic order) as the original orbital $\chi_\mu$. Also, the $h(|\mathbf{r}'-\mathbf{r}|; R)$ of eq 52 is a linear combination of Gaussians, and the convolution of a Gaussian-type orbital by a spherical Gaussian is
\begin{align}
    \int \dd[3]\mathbf{r}' e^{-\beta(\mathbf{r}'-\mathbf{r})^2} \chi_\nu(\mathbf{r}') &= h_{An\ell}(|\mathbf{r}_A|; \beta) Y_{\ell m}(\hat{\mathbf{r}}_A) \label{eq:gto} \\
    h_{An\ell}(r; \beta) &= r^\ell \sum_j D_{An\ell,j}(\beta) \exp\left(-\frac{\alpha_j\beta}{\alpha_j+\beta} r^2\right) \\
    D_{An\ell,j}(\beta) &= \left(\frac{\pi}{\beta}\right)^{3/2} \left(\frac{\beta}{\alpha_j+\beta}\right)^{3/2+l/2} C_{An\ell,j}
\end{align}
These convolutions are also Gaussian-type orbitals, so they can be evaluated using the existing atomic orbital evaluation code in PySCF with some minor modifications. The evaluation of eq~\ref{eq:gamma_conv} can be accelerated by computing the radial part $h_{An\ell}(|\mathbf{r}_A|; \beta)$ and the spherical harmonics $Y_{\ell m}$ separately, since the radial part does not depend on $m$ and the spherical harmonics do not depend on $\beta$.

\subsection{Periodic Systems (with Pseudopotentials)}\label{sec:si_periodic}

For a periodic system with pseudopotentials, the orbitals and density contributions are sufficiently smooth to use a uniform grid and therefore take advantage of the Fast Fourier Transform (FFT) to efficiently convert functions between real and reciprocal space.

The first step is to evaluate the atomic orbitals in reciprocal space, eq~\ref{eq:gto_fft}. Next we perform the convolutions of these orbitals by $h(\mathbf{r})$ in reciprocal space:
\begin{align}
    \int \dd[3]\mathbf{r}' h(|\mathbf{r}'-\mathbf{r}|; R) \chi_\nu(\mathbf{r}') &= \mathcal{F}\left\{ \tilde{h}(\mathbf{k}; R) \tilde{\chi}_\nu(\mathbf{k}) \right\}(\mathbf{r}) \label{eq:convolved_aos_recip} \\
    \nabla \left(\int \dd[3]\mathbf{r}' h(|\mathbf{r}'-\mathbf{r}|; R) \chi_\nu(\mathbf{r}')\right) &= \mathcal{F}\left\{ i\mathbf{k} \tilde{h}(\mathbf{k}; R) \tilde{\chi}_\nu(\mathbf{k}) \right\}(\mathbf{r}) \label{eq:convolved_daos_recip} \\
\end{align}
where
\begin{equation}
    \tilde{h}(\mathbf{k}; R) = \mathcal{F} \left\{ h(|\mathbf{r}|; R) \right\}(\mathbf{r})
\end{equation}
Using eqs~\ref{eq:convolved_aos_recip} and~\ref{eq:convolved_daos_recip}, eqs~\ref{eq:rho0_ao} and~\ref{eq:rho1_ao} can be evaluated from the orbital density matrix $\mathbf{P}$. Alternatively, if one has access to the molecular orbitals, one can evaluate
\begin{equation}
    \int \dd[3]\mathbf{r}' h(|\mathbf{r}'-\mathbf{r}|; R) \phi_i(\mathbf{r}')
\end{equation}
and its gradient analogously to eqs~\ref{eq:convolved_aos_recip} and~\ref{eq:convolved_daos_recip}, then substitute these integrals to compute eqs~\ref{eq:rho0_mo} and~\ref{eq:rho1_mo}.

\section{Plots of Polaron Levels with Respect to Fraction of CIDER Exchange}

The exact exchange-correlation functional, as discussed in the main text, is piecewise linear with respect to the number of electrons. As a condition of this property, for an integer electron number $N$, the conduction band minimum energy level (CB) of the $N$-electron system and the valence band maximum energy level (VB) of the $N+1$-electron system are the same:
\begin{equation}
    \epsilon_\text{CB}^\text{exact}(N) = \epsilon_\text{VB}^\text{exact}(N+1) \label{eq:eigval_relationship_exact}
\end{equation}
Due to self-interaction error, semilocal (SL) functionals are in general convex with respect to electron number, which means
\begin{equation}
    \epsilon_\text{CB}^\text{SL}(N) > \epsilon_\text{VB}^\text{SL}(N+1)
\end{equation}
Self-interaction correction (SIC) terms, like the $U$ term in DFT+$U$, the exact or CIDER exchange energy, etc., are in general concave, which means
\begin{equation}
    \epsilon_\text{CB}^\text{SIC}(N) < \epsilon_\text{VB}^\text{SIC}(N+1)
\end{equation}
Therefore, there should be some fraction of a given self-interaction correction that can be added to a semilocal functional such that the exact condition of eq~\ref{eq:eigval_relationship_exact} is obeyed. For CIDER exchange, we use the functional form
\begin{equation}
    E_\text{xc} = E_\text{xc}^\text{PBE} + \alpha (E_\text{x}^\text{CIDER} - E_\text{x}^\text{PBE})
\end{equation}
so $E_\text{x}^\text{CIDER} - E_\text{x}^\text{PBE}$ serves as the self-interaction correction term.

\begin{figure}
    \centering
    \includegraphics[width=0.6\textwidth]{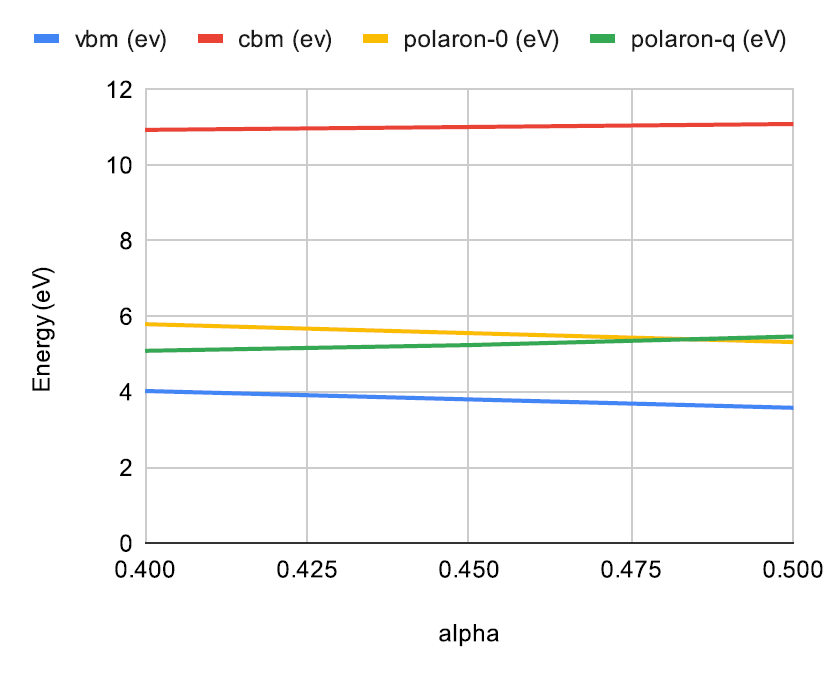}
    \caption{Energy levels of the valence band maximum (VBM), conduction band minimum (CBM), polaron level of the uncharged distorted system (polaron-0), and the polaron level of the charged distorted system (polaron-q) for \ce{MgO}.}
    \label{fig:mgo_levels}
\end{figure}

\begin{figure}
    \centering
    \includegraphics[width=0.6\textwidth]{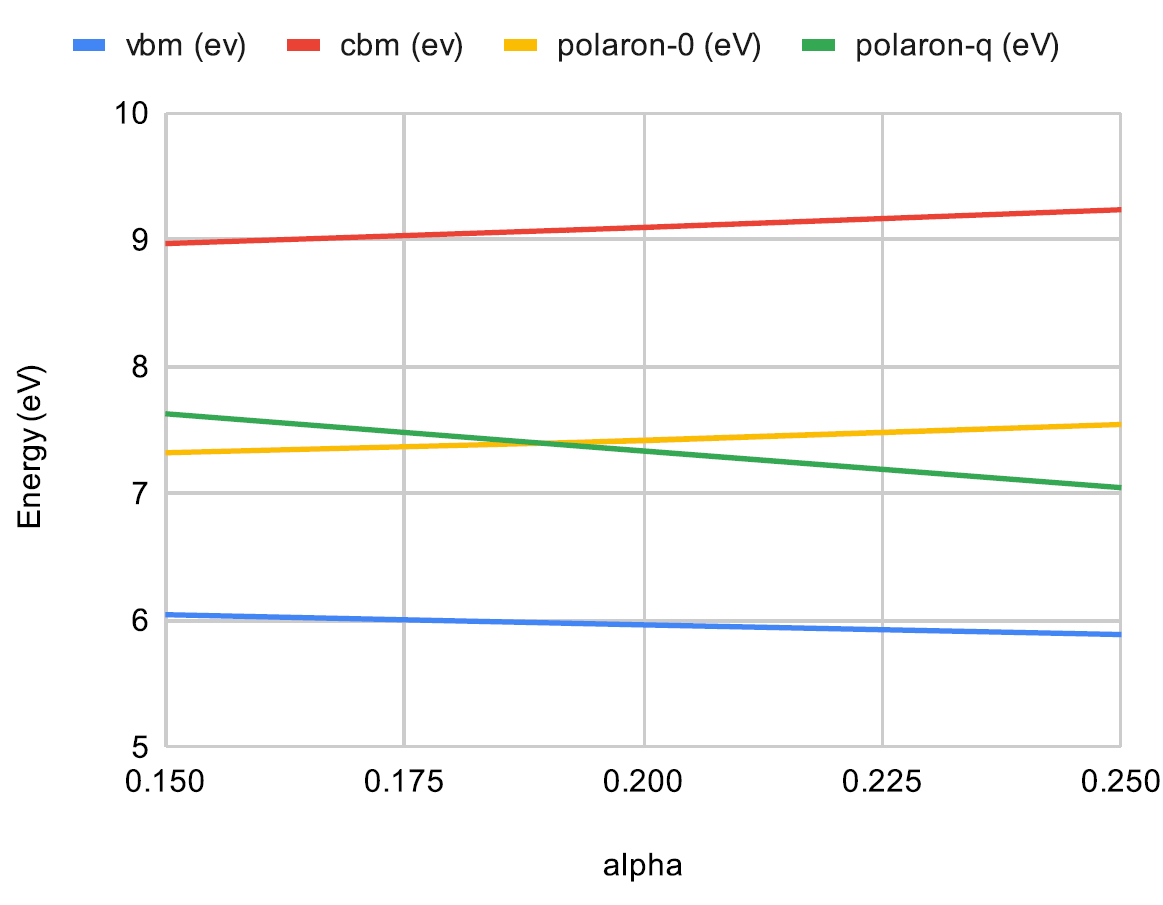}
    \caption{Energy levels of the valence band maximum (VBM), conduction band minimum (CBM), polaron level of the uncharged distorted system (polaron-0), and the polaron level of the charged distorted system (polaron-q) for \ce{BiVO4}.}
    \label{fig:bivo4_levels}
\end{figure}

For a hole polaron, $\epsilon_\text{CB}(N)$ is the lowest unoccupied eigenvalue of the charged system $\epsilon_\text{polaron}(q)$, and $\epsilon_\text{VB}(N+1)$ is the highest occupied eigenvalue of the uncharged system $\epsilon_\text{polaron}(0)$, both computed at the equilibrium geometry of the charged system. For an electron polaron, $\epsilon_\text{CB}(N)$ is the lowest unoccupied eigenvalue of the uncharged system $\epsilon_\text{polaron}(0)$, and $\epsilon_\text{VB}(N+1)$ is the highest occupied eigenvalue of the charged system $\epsilon_\text{p}(q)$. If CIDER serves as a good proxy for exact exchange and therefore as a good self-interaction correction, $\epsilon_\text{polaron}(0)$ and $\epsilon_\text{polaron}(q)$ should approach and then cross each other as $\alpha$ is increased. Figures~\ref{fig:mgo_levels} and~\ref{fig:bivo4_levels} show that this is indeed the case for the hole polaron in \ce{MgO} and electron polaron in \ce{BiVO4}, respectively. In \ce{MgO}, the uncharged polaron level tracks with the valence band and decreases in energy as $\alpha$ increases, eventually crossing the charged polaron level, which increases in energy, at $\alpha\approx 0.484$. In \ce{BiVO4}, the uncharged level increases in energy while the charged level decreases in energy, and the levels cross as $\alpha\approx 0.192$. In both cases, the behavior of the polaron levels computed with CIDER matches the expected behavior found with other methods~\cite{falletta_polarons_2022,falletta_hubbard_2022,falletta2024JAP}.

\begin{figure}
    \centering
    \includegraphics[width=0.22\textwidth]{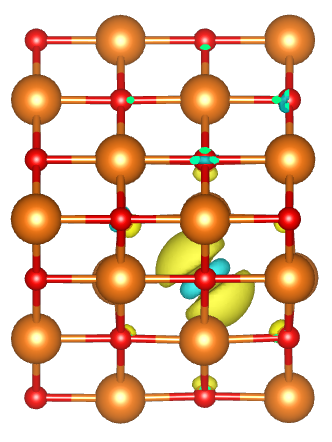}
    \includegraphics[width=0.27\textwidth]{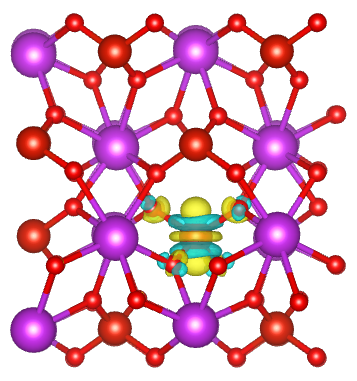}
    \caption{Charge distributions of the hole polaron in \ce{MgO} (left) and of the electron polaron in \ce{BiVO4} (right), obtained using CIDER24X-ne($\alpha$) with $\alpha$ from Table 2. Visualization performed using Vesta~\cite{Momma2008}.}
    \label{fig:polaron_chg_loc}
\end{figure}

\bibliography{references}